\documentclass[11pt]{article}
\usepackage{amsmath,amsfonts}
\def \beq  {\begin{equation}}
\def \eeq  {\end{equation}}
\def \beqar {\begin{eqnarray}}
\def \eeqar {\end{eqnarray}}
\def\sqr#1#2{{\vcenter{\vbox{\hrule height.#2pt
\hbox{\vrule width.#2pt height#1pt \kern#1pt
\vrule width.#2pt}\hrule height.#2pt}}}}

\def\la {{\langle}}
\def\ra {{\rangle}}

\def\Tr {{\rm Tr}}
\def \tr {{\rm tr}}

\def\bxi {{\bar \xi}}

\def\del {\partial}

\def\D {{\cal D}}
\def\bz {{\bar{z}}}

\def\half{\textstyle{1\over 2}}

\def \onethird {{1\over 3}}

\begin{document}
%%%%%%%%%%%%%%%%%%%%%%%%%%%%%%%%%%%%%%%%%%%%%%%
%\fontfamily{pnb}\fontsize{12pt}{16pt}\selectfont
%\fontfamily{pzc}\fontsize{14pt}{16pt}\selectfont
%\fontfamily{pbk}\fontsize{12pt}{16pt}\selectfont
%\fontfamily{cmr}\fontsize{12pt}{16pt}\selectfont
%\fontfamily{phv}\fontshape{ro}\fontsize{11pt}{14pt}\selectfont
%\fontfamily{ptm}\fontseries{m}\fontshape{r}\fontsize{12pt}{16pt}\selectfont
%\fontfamily{pnc}\fontseries{m}\fontshape{r}\fontsize{11pt}{14pt}\selectfont
%\usefont{T1}{phv}{m}{it}
%%%%%%%%%%%%%%%%%%%%%%%%%%%%%%%%%%%%%%%%%%%%%%%
\def \CMP {{ Commun. Math. Phys.}}
\def \PRL {{ Phys. Rev. Lett.}}
\def \PL {{Phys. Lett.}}
\def \NPBProc {{ Nucl. Phys. B (Proc. Suppl.)}}
\def \NP {{ Nucl. Phys.}}
\def \RMP {{ Rev. Mod. Phys.}}
\def \JGP {{ J. Geom. Phys.}}
\def \CQG {{ Class. Quant. Grav.}}
\def \MPL {{Mod. Phys. Lett.}}
\def \IJMP {{ Int. J. Mod. Phys.}}
\def \JHEP {{ JHEP}}
\def \PR {{Phys. Rev.}}
\def \JMP {{J. Math. Phys.}}
%%%%%%%%%%%%%%%%%%%%%%%%%%%%%%%%%%%%%%%%%%%%%%%
%%%%%%%%%%%%%%%%%%%%%%%%%%%%%%%%%%%%%%%%%%%%%%%
\begin{titlepage}
\null\vspace{-62pt} \pagestyle{empty}
\begin{center}
\rightline{CCNY-HEP-06/4}
\rightline{April 2006}
\vspace{1truein} {\Large\bfseries
The Matrix Chern-Simons One-form as a Universal}\\
\vskip .2in
{\Large\bfseries  Chern-Simons Theory}
\vskip .2in\noindent

%%%%%%%%%%%%%%%%%%%%%%%%%%%%%%%%%%%%%%%%%%%%%%%%%
\vspace{.5in}
{\bf\large V. P. NAIR}\\
\vspace{.15in}{\itshape Physics Department\\
City College of the CUNY\\  
New York, NY 10031}\\
E-mail:
\fontfamily{cmtt}\fontsize{11pt}{15pt}\selectfont vpn@sci.ccny.cuny.edu

\fontfamily{cmr}\fontsize{11pt}{15pt}\selectfont
\vspace{.4in}
%\vspace{1.5in}%\vspace{0.3in}
%%%%%%%%%%%%%%%%%%%%%%%%%%%%%%%%%%%%%%%%%%%%%%%%%%%%%%%%%%%%
\centerline{\large\bf Abstract}
\end{center}
We consider different large ${\cal N}$ limits of the one-dimensional Chern-Simons action
$i\int dt~ \Tr (\del_0 +A_0)$ where $A_0$ is an ${\cal N}\times{\cal N}$ antihermitian
matrix. The Hilbert space on which $A_0$ acts as a linear transformation is taken as the quantization of a $2k$-dimensional phase space ${\cal M}$ with different 
gauge field backgrounds. For slowly varying fields, the large ${\cal N}$ limit of the 
one-dimensional CS action is equal to the $(2k+1)$-dimensional CS theory
on ${\cal M}\times {\bf R}$. Different large ${\cal N}$ limits are parametrized by the gauge fields and the dimension $2k$. The result is related to the bulk action for quantum Hall droplets in higher dimensions. Since the isometries of ${\cal M}$ are gauged, this has implications for gravity on fuzzy spaces. This is also briefly discussed.
\end{titlepage}
%%%%%%%%%%%%%%%%%%%%%%%%%%%%%%%%%%%%%%%%%%%%%%%%%%%%%%
\pagestyle{plain} \setcounter{page}{2}

\section{Introduction}

Over the last few years there has been a lot of research interest in noncommutative and fuzzy spaces \cite{general}.
Such spaces can arise as solutions in string and $M$-theories.
For example, in the matrix model version of $M$-theory, noncommutative branes can be obtained
as $(N\times N)$-matrix configurations whose large $N$ limits tend to smooth manifolds.
Fluctuations of such branes are described by gauge theories and, naturally for this reason, there has been a large number of papers on these spaces.
For some compact manifolds of finite volume, it is possible to obtain noncommutative versions which are described by finite dimensional matrices \cite{bal1}. Being finite dimensional, these matrices
give a fuzzy version of the manifold. Fields on such spaces have a finite number of modes, and therefore, the fuzzy version of a manifold may also be thought of as a regularization for the  manifold itself and for field theories defined on it.
As a regularization, fuzzification has the advantage of preserving various symmetries which we expect for the smooth manifold. In the context of branes, matrix models (or fuzzy versions)
have even become a fairly standard method of analysis \cite{mald}.
Some other advantages of fuzzification, such as application to the fermion doubling problem, have also been investigated \cite{bal2}.

One interesting class of noncommutative field theories which has been extensively investigated is the Chern-Simons theory \cite{poly}. Properties of these theories on flat noncommutative spaces are fairly well understood by now. Their formulation on fuzzy spaces, i.e., in terms 
of finite dimensional matrices, has been a little tricky. This has been achieved for a number of two-dimensional spaces, although a general formulation is not yet possible. In this article, we will investigate the one-dimensional Chern-Simons theory which is given by the action
$S =i \int dt~ \Tr (D_0)$, where $D_0$ is some $({\cal N} \times {\cal N})$ antihermitian
matrix . We consider different large ${\cal N}$ limits of this theory.
For the purpose of the large ${\cal N}$ analysis, let ${\cal H}$ denote the vector space on which $D_0$ is a linear transformation. We may regard ${\cal H}$ as the quantization
of the phase space $S^2 = SU(2)/U(1)$, where the symplectic form is $\omega = -in \omega_K$, $\omega_K$ being the K\"ahler form on $S^2$. (${\cal N}$ will be a function of $n$.)
Taking $n$ large in this way defines a specific large ${\cal N}$ limit.
Notice that $\omega$ is a background $U(1)$ field on $S^2$, and so, we could also consider a deformation of this situation where $\omega = -in \omega_K +F$,
where $F$ is topologically trivial (so that the dimension of ${\cal H}$ is not changed).
The large $n$ limit now gives $S^2$ with a different choice of background field on it.
One could also consider ${\cal H}$ as the quantization of, say, ${\bf CP}^k = SU(k+1)/U(k)$,
with a suitable choice of symplectic form (with the dimensions of ${\cal H}$'s matching
dimensions of a class of $SU(k+1)$ irreducible representations).
This would give yet another set of large ${\cal N}$ limits. In a gradient expansion where the fields (like $F$) are slowly varying, we can then show that, as ${\cal N}$ becomes large,
\beq
i \int dt~ \Tr (D_0) = S_{CS} (a +A) ~+~\cdots
\label{intro1}
\eeq
where $a$ is the potential for the symplectic form and $A$ is any background field on the space. $S^{(2k+1)}_{CS}$ is the $(2k+1)$-dimensional Chern-Simons form and the omitted terms are smaller by powers of $n$ or involve higher gradients of fields.
The choice of $a$, $A$ parametrize the different large ${\cal N}$ limits which give the right hand side of (\ref{intro1}).  This equation shows that we may consider the one-dimensional
Chern-Simons form as a `universal' Chern-Simons theory.
We may, for example, use it as a regularized version of the CS theory on $S^2 \times {\bf R}$, or on ${\bf CP}^2 \times {\bf R}$, or on various other spaces, the particular manifolds being recovered as specific large $n$ limits.

We now come to the motivation for this result. A first application is to the quantum Hall effect in higher dimensions which has been extensively studied over the last few years
\cite{ZH, KN, other}.
The lowest Landau level for the Landau problem (of a charged particle in magnetic field)
on an even dimensional space
${\cal M}$ defines a Hilbert space ${\cal H}$ which is the quantization of ${\cal M}$ with a symplectic form which is the background magnetic field \cite{KN}.
The lowest Landau level is thus a model for the fuzzy version of ${\cal M}$ and dynamics
confined to the lowest Landau level is a matrix model to which all of the analysis 
mentioned above can be applied \cite{KNR}.
Our result then leads to the computation of the bulk action for quantum Hall droplets.
The bulk action, with fluctuations of the gauge fields, it should be mentioned, has previously been obtained by other methods.
In reference \cite{karabali}, the bulk action is obtained by requiring that the gauge transformation of $D_0$ is induced by the gauge transformation
of the background fields. This leads to an elegant calculation of the
bulk and boundary actions. The anomaly cancellation between the bulk and boundary actions is also demonstrated. (The boundary actions, without gauge field fluctuations, have been obtained previously in \cite{KN} for arbitrary dimensions. The case of the two-dimensional flat
space has been analyzed along similar lines in \cite{sakita}.)
Needless to say, the bulk action derived by the methods in this paper
agrees with the results in \cite{karabali}.

The variation of the background gauge fields can also be interpreted as gravitational perturbations of the manifold since the isometry groups are being gauged.
Thus gravity on fuzzy spaces is another natural application of our analysis.
In fact, we present arguments that $i\int dt~ \Tr (D_0)$ is the natural action for gravity
on fuzzy spaces. This is outlined here and explained more fully in a separate article \cite{nair}.
In the large ${\cal N}$ limit of smooth manifolds, what is obtained is then Chern-Simons gravity. It is interesting to point out at this stage that there are indications that $M$-theory can lead to CS gravity
\cite{horava}.
We expect that our results do have some implications for the matrix analysis of $M$-theory.

This paper is organized as follows. In section 2, we give a resum\'e of some basic results
on the quantum Hall effect on ${\bf CP}^k$ and the fuzzy version of
${\bf CP}^k$ which are needed for the subsequent analysis. 
Section 3 is a brief aside showing how our results, at least for the Abelian case, 
can be obtained by a simple phase space analysis without the use of
matrices.
In section 4, we give the main arguments leading to the result (\ref{intro1}). Since this is a long technical section, we first show how the result (\ref{intro1}) emerges for $S^2 \times {\bf R}$.
This will also show how we may interpret (\ref{intro1}) as a 
fuzzy version of the CS action on spaces of different topologies.
The following subsections of section 4 derive the general result. In section 5, we discuss applying our result to the quantum Hall effect and to fuzzy gravity.

\section{Resum\'e of Hall effect on ${\bf CP}^k$ and fuzzy ${\bf CP}^k$}

We begin by considering Hall effect on ${\bf CP}^k = SU(k+1)/U(k)$, concentrating on the lowest Landau level. The states corresponding to the lowest Landau level can also be considered as the (finite-dimensional) Hilbert space ${\cal H}$ which is part of the definition of fuzzy ${\bf CP}^k$. With this re-interpretation, our results can be applied to matrix models and to fuzzy spaces.

Let $g$ denote a $(k+1)\times (k+1)$ matrix corresponding to a general element
of $SU(k+1)$ in the fundamental representation. Further, let $t_a$,
$a = 1, 2, \cdots, k^2+2k$, be a set of hermitian matrices which form a basis of the Lie algebra of $SU(k+1)$, again in the fundamental representation. These are taken to obey
\beq
[ t_a, t_b ] = i f_{abc} t_c , \hskip 1in \Tr (t_a t_b )= \half \delta_{ab}
\label{res1}
\eeq
$f_{abc}$ are the structure constants of $SU(k+1)$ in this basis.
On $g$, we can define left and right translation operators by
\beq
L_a ~g = t_a ~g, \hskip 1in R_a ~g = g~t_a
\label{res2}
\eeq
It is convenient to split these into the $R_{k^2+2k}$ which is the $U(1)$ generator
in $U(k)\subset SU(k+1)$, $R_j$, $j= 1, 2, \cdots, k^2-1$, which are $SU(k)$ generators
and $R_{\pm i}$, $i=1, 2, \cdots, k$ which are in the complement of
$\underline {U(k)}$ in the Lie algebra $\underline{SU(k+1)}$. (A similar splitting can be made for the left translations, but will not be necessary for what follows.)

Since ${\bf CP}^k = SU(k+1) /U(k)$, we can choose constant background magnetic fields
which take values $\underline{U(k)}$, the Lie algebra of $U(k)$; in other words, we can have
$\underline{SU(k)}$-valued and $\underline{U(1)}$-valued fields. These will be proportional to the Riemann curvatures of ${\bf CP}^k$. Both possibilities have been considered and explored in some detail before. Here it is sufficient to consider the case of $\underline{U(1)}$ background.
In this case, the wave functions must obey 
\beqar
R_j ~\Psi (g)&=&0, ~~~~~~~~~~~~~~~j=1,\cdots , k^2 -1 \nonumber \\
R_{k^2 + 2k} ~\Psi (g)&=& nk {1 \over {\sqrt{2k(k+1)}}}~\Psi (g)
\label{res3}
\eeqar
where $n$ is an integer characterizing the strength of the field. (The integrality of $n$
can be understood via standard arguments similar to the Dirac quantization for monopoles.)
For the lowest Landau level, we have an additional condition
\footnote{In references \cite{KN}, we used $R_{k^2+2k}\Psi = -nk/\sqrt{2k(k+1)}~\Psi$. The symbols were then defined using $g^*$ and $g^T$. It is simpler to use $g,~g^\dagger$ in the definition of the symbol. Since $\left(e^{it\cdot \theta} g \right)^* = e^{i (-t^T)\cdot \theta} g^*$,
we can use $g, ~g^\dagger$ for the symbol if the $t$-matrices in this paper are taken as 
$- t^T$ of the usual fundamental representation for $SU(k+1)$ or the basis used
in \cite{KN}. Thus, for example, for $SU(3)$,
$t_a$ in this paper are $t_a = -\half \lambda_a^T$ in terms of the Gell-Mann matrices
$\{ \lambda_a\}$.
This is equivalent to using the conjugate representation, $t \rightarrow -t^T$ being the
automorphism of the Lie algebra corresponding to conjugation.}
\beq
R_{+i}  ~\Psi (g) =0
\label{res4}
\eeq
As discussed in \cite{KN}, the wave functions obeying the conditions
(\ref{res3}) and (\ref{res4}) are given in terms of the Wigner ${\cal D}$-functions of
$SU(k+1)$. These can be defined as
\beq
{\cal D}^{(J)}_{pq}(g) \equiv \la J, p \vert~ {\hat g}~ \vert J, q\ra
\label{res5}
\eeq
Evidently, these are the representatives of $g$ in the representation $J$ and may be thought of as the matrix elements of a general operator version of $g$, namely ${\hat g}$, with the states of the representation $J$.
The conditions (\ref{res3}) and (\ref{res4}) tell us that, for the case at hand, we need
${\cal D}^{(n)}_{m, n} (g) = \la n, m\vert {\hat g} \vert n, n\ra$, corresponding to the symmetric rank $n$ representation, denoted by $(n)$, and the state on the right 
is chosen to be the highest weight state $\vert w\ra = \vert n, n\ra$, which is an $SU(k)$-singlet
and has the charge $nk/\sqrt{2k(k+1)}$ for $R_{k^2+2k}$.
Being the highest weight state, we also get the condition (\ref{res4}).
The left index $m$ can take $N$ values, where $N$ is the dimension of the symmetric
rank $n$ representation; explicitly
\beq
N = {\rm dim} ~J= {(n+k)! \over n! ~k!}
\label{res6}
\eeq
The properly normalized wave functions are
\beq
\Psi_m (g) = \sqrt{N} ~ {\cal D}^{(n)}_{m, n} (g) 
\label{res7}
\eeq
These are normalized by virtue of the orthogonality theorem
\beq
\int d\mu (g) ~ {\cal D}^{*(n)}_{m, n} (g)  {\cal D}^{(n)}_{m', n} (g) 
= {\delta_{mm'}\over N}\label{res8}
\eeq
Notice that the wave functions carry a representation of the left action of $SU(k+1)$; this corresponds to the freedom of magnetic translations in the Hall effect interpretation.

We can introduce a local coordinate description of these as follows.
Let $g_{\alpha, k+1}\equiv u_{\alpha}$, with $\alpha = 1, 2, \cdots, k+1$.
In terms of the $u_\alpha$, 
$\D^{(n)}_{m , n} \sim u_{\alpha_1} u_{\alpha_2}\cdots u_{\alpha_n}$.
Since $u^*\cdot u =1$, we can parametrize them as
\beqar
u_i &=& {\xi_i \over \sqrt{1+\bxi \cdot \xi }}~~~~~~~~~~~~~i=1,\cdots , k 
\nonumber \\
u_{k+1} &=& {1\over \sqrt{1+\bxi \cdot \xi }}
\label{res9}
\eeqar
The Wigner ${\cal D}$-functions are now
\beqar
{\cal D}^{(n)}_{m, n}(g) &=& \left[ {n! \over i_1! i_2! ...i_k!
(n-s)!}\right]^{\half} ~ {\xi_1^{i_1} \xi_2^{i_2}\cdots \xi_k^{i_k}\over
(1+\bxi \cdot \xi )^{n/2} }\nonumber\\
s &=& i_1 + i_2 + \cdots + i_k \label{res10}
\eeqar

The complex projective space ${\bf CP}^k$ is a K\"ahler manifold; the K\"ahler form
may be written in terms of the local coordinates as
\beq
\omega_K = i \left[ {d \xi ^i ~d\bxi^i \over (1+\bxi\cdot \xi)}
- { \bxi^i d\xi^i ~ \xi^j d\bxi^j \over (1+\bxi\cdot \xi )^2}
\right]
\label{res11}
\eeq
The background field we have chosen is given by
$-in\omega_K$ (in an antihermitian basis). The states are obtained by quantization of the symplectic form
$\omega =  -in \omega_K$. Thus the lowest Landau level wave functions correspond to sections of
the  line bundle with curvature $-in \omega_K$.

The volume of ${\bf CP}^k$ is given by
\beq
d\mu ({\bf CP}^k) = \left({ \omega_K \over 2\pi}\right)^k =
{k! \over \pi^k} {d^2\xi^1 d^2\xi^2\cdots d^2\xi^k \over (1+\bxi\cdot \xi )^{k+1}}
\label{res12}
\eeq
In real coordinates this can be written as
\beq
d\mu ({\bf CP}^k) ={1\over (2\pi)^k} \left[ {1\over 2} (\omega_{K})_{\mu\nu} dx^\mu dx^\nu\right]^k
=
{1\over (4\pi)^k}  2^k k! \sqrt{\det \omega_K} ~d^{2k}x
\label{res13}
\eeq
In equation (\ref{res12}), the quantity $d^2\xi = d\xi d\bxi $ is equal to
$d^2x$ in real coordinates.

The inner product for the wave functions can also be written in terms of this measure
and reads
\beq
\la \Psi \vert \Psi' \ra = \int d\mu ({\bf CP}^k) ~\Psi^* ~\Psi'
\label{res14}
\eeq

The states in the lowest Landau level 
are elements of a finite-dimensional Hilbert space ${\cal H}_N$. This space is
what is obtained by quantizing the symplectic form $-in \omega_K$.
The algebra of matrices which are linear transformations on ${\cal H}_N$ tend to
the algebra of functions on ${\bf CP}^k$ as $N$ becomes large.
We may therefore identify the space ${\cal H}_N$ as defining a fuzzy 
version of ${\bf CP}^k$. (There is a natural choice of Laplace operator
on the set of matrices which will complete the required triple of
algebra, Hilbert space and Laplacian.)

The notion of the symbol is useful in analyzing the large $N$ limit which leads to
the continuous ${\bf CP}^k$. 
If ${\hat A}$ is a matrix, taken as a linear transformation on ${\cal H}_N$, then the symbol
corresponding to ${\hat A}$ is given by
\beq
({\hat A}) = A(g) =  \sum_{ms}\D^{(n)*}_{m, n}
(g) A_{ms} \D^{(n)}_{s, n}(g) = \la w\vert ~ {\hat g}^{-1} {\hat A}~ {\hat g}~\vert w\ra
\label{res15}
\eeq

The symbol
of ${\hat A}$ coincides with the expectation value of ${\hat A}$
in the large $n$ limit.
The symbol corresponding to the product of two matrices is given by the star product
of the symbols of the matrices. It can be written as
\beqar
({\hat A} {\hat  B})\equiv ({\hat A})* ({\hat B})&=& \sum_s (-1)^s \left[ {(n-s)! \over n! ~s!}\right]
\sum_{\sum_k i_k=s}^n~
{s! \over i_1! i_2! \cdots i_k!}~{\hat R}_{-1}^{i_1} {\hat R}_{-2}^{i_2}
\cdots {\hat R}_{-k}^{i_k} A(g)\nonumber\\
&&\hskip 2.5in\times~
{\hat R}_{+1}^{i_1} {\hat R}_{+2}^{i_2}\cdots {\hat R}_{+k}^{i_k}
 B(g)\nonumber\\
 &=& A B ~-{1\over n} {\hat R}_{-i} A ~{\hat R}_{+i} B ~+~{\cal O}(1/n^2)
 \label{res16}
\eeqar

The trace of a matrix can be related to the integral over the phase space of the
symbol of the matrix,
\beq
\Tr~ {\hat A} = \sum_{m} A_{mm} = N~\int d\mu (g) ~A(g)
\label{res17}
\eeq

Let $T_a$ be matrices corresponding to $t_a$ in the representation $J$; they are
$N\times N$-matrices, obeying the commutation rules in (\ref{res1}).
The symbol for the action of $T_a$ on a matrix ${\hat A}$ can be worked out as 
follows.
\beqar
(T_a {\hat A}) &=& \la w\vert ~{\hat g}^{-1}~ T_a~ {\hat A}~ {\hat g} ~\vert w\ra
\nonumber\\
&=& S_{a b} ~\la w\vert ~T_b ~{\hat g}^{-1} ~{\hat A}~ {\hat g}~\vert w\ra\nonumber\\
&=& {nk\over \sqrt{2k(k+1)}} S_{a k^2+2k} ~A(g) + \sum_{i=1}^{2k}S_{a i} \la w\vert T_i~ {\hat g}^{-1} {\hat A} {\hat g}\vert w\ra\nonumber\\
&=& \left[ {nk\over \sqrt{2k(k+1)}} S_{a k^2+2k} - {1\over 2} \sum^k ~S_{a -i}  ~ R_{+i}
\right] A(g)
\label{res18}
\eeqar
where $S_{a b} = 2~ \Tr (g^{-1} t_a g t_b )$ and we have also used
the fact that $T_+ \vert w\ra =0$. Further, $R_a g = g t_a$, and $R_a g^{-1}
= - t_a g^{-1}$. In a similar way,
\beq
({\hat A}T_a )= \left[  {nk\over \sqrt{2k(k+1)}} S_{a k^2+2k} + {1\over 2}
\sum^k S_{a +i}~R_{-i}
\right] A(g)
\label{res19}
\eeq
These results also show that
\beq
([-iT_a , {\hat A}]) = \sum_{i=1}^{2k} S_{a i} ~i R_i ~A(g)
\label{res20}
\eeq
The summation is over the basis of the coset $\underline{SU(k+1)} - \underline{U(k)}$.
Notice also that the large $n$ limit of $T_a$ are given by the $S_{a k^2 +2k}$ 
which may be interpreted as giving the Cartesian coordinates for embedding 
of ${\bf CP}^k$
in ${\bf R}^{k^2 +2k}$.

It is also useful to describe fuzzy ${\bf CP}^k$, before we take the large $n$ limit,
directly in terms of embedding in ${\bf R}^{k^2 +2k}$. For this we start with
$k^2 +2k$ hermitian matrices $X_a$ which are of dimension
$(N\times N)$ .
We take $N$ to be of the form (\ref{res6}) for some integer $n$.
The embedding conditions are then given by \cite{bal3}
\beqar
X_a X_a &=& {nk(n+k+1) \over 2(k+1)}~\equiv C_n\nonumber\\
d_{abc} X_b X_c &=& (k-1) {(2n +k +1) \over 4 (k+1)}~ X_a ~\equiv \alpha_n ~X_a
\label{res21}
\eeqar

Consider the $SU(k+1)$-generators $T_a$ in the symmetric representation
of rank $n$. They may be written as
$T_a = a^\dagger_\alpha (t_a)_{\alpha\beta} a_\beta \equiv a^\dagger t_a a$, for bosonic annihilation-creation
operators $a_\beta, a^\dagger_\alpha$, $\alpha, \beta = 1, ..., (k+1)$.
By using completeness relations, it is easy to prove that these obey representation-dependent identities which are identical to (\ref{res21}) with $T_a$ replacing $X_a$. In other words,
the matrices $T_a$ in the symmetric rank $n$ representation of $SU(k+1)$
give a solution of the embedding conditions (\ref{res21}) via $X_a = T_a
= a^\dagger t_a a$. In equation (\ref{res21}), $C_n$  is the quadratic Casimir operator
and $\alpha_n$ is another invariant related to the properties of the $d_{abc}$-symbol.

We may rewrite the conditions (\ref{res21}) in terms of $-iT_a$ as
\beqar
(-iT_a) (-iT_a) &=& - C_n \nonumber\\
d_{abc} (-iT_b ) (-iT_c) &=& -i \alpha_n  (-iT_a)\label{res22}
\eeqar
 
\section{The effective bulk action for the $U(1)$ background}

We consider Hall droplets on a ${\bf CP}^k$ with symplectic two-form 
$\omega = \half \omega_{ij} d\xi^i \wedge d\xi^j$. As mentioned in the previous section,
the lowest Landau levels are described by the quantization of the space with this symplectic form.
The background magnetic field is given by $\omega$ itself, with a potential $a(\xi )$ such that
$\omega = da$.  We now consider the addition of a $U(1)$ magnetic field described by a gauge potential $A$; in other words the total potential is now $a +A$.
The new dynamics is given by the symplectic form $\omega +F$. Let $H= V$ be the original Hamiltonian. The new Hamiltonian is $H= V +A_0$, where $A_0$ is the time-component
of the gauge potential.
We consider the case where $\omega +F$ has the same phase volume, so that the number of states in the Hilbert space, the lowest Landau level, is not changed.
We can now pose the following problem: Describe the dynamical system of $(\omega +F, H= V+A_0 )$ as $(\omega , H= {\cal A})$, where we use the old symplectic form but a different Hamiltonian. At least for the $U(1)$ case, this can be accomplished by choosing a new set of variables in phase space.
We write $\xi = v -w$, where $v$'s are coordinates on ${\bf CP}^k$ and $w$ is viewed as a series in powers of $F$ and derivatives of $A, A_0$. To accomplish the change to
using $\omega$ as the symplectic form, we need
\beq
(\omega + F )\biggr]_{\xi} = ~\omega \biggr]_v
\label{ab1}
\eeq
The change $\xi \rightarrow v$ is a coordinate transformation which is not canonical.
The strategy is to find $w$ as a series  and then the Hamiltonian is given by
\beq
H \equiv {\cal A} = (V+A_0)\biggr]_{v-w}
\label{ab2}
\eeq
Rather than implementing ({\ref {ab1}) at the level of $\omega$'s,  we can use the potentials and write the equivalent version of (\ref{ab1}) as
\beq
(a +A)\biggr]_{v-w} ~-~ a\biggr]_v = d \theta
\label{ab3}
\eeq
where $\theta$  is arbitrary. It can be considered as a gauge transformation of the gauge potentials; evidently it does not contribute to the symplectic forms. Written out in coordinate
notation, this equation becomes
\beq
[ a_i (v-w) - a_i (v) ] - a_j (v-w) \del_i w^j + A_i (v-w) - A_j(v-w) \del_i w^j \approx 0
\label{ab4}
\eeq
where the weak equality sign $\approx$ indicates equivalence up to a gauge transformation.
We now write $w = \lambda w_1 + \lambda^2 w_2 +\cdots$, where $\lambda$ is a parameter introduced to keep track of various terms of the same order; eventually it is set to one.
The potential $V$ is taken to be of order $\lambda^0$ and the fluctuation $A$ is taken to be of order one, thus $A\rightarrow \lambda A$.
The requirement (\ref{ab4}) can be written, for the first two powers of $\lambda$, as
\beqar
w_1 \cdot \del a_i + a_j \del_i w_1^j - A_i &\approx& 0\label{ab5}\\
w_2\cdot \del a_i + a_j \del_i w_2^j + w_1 \cdot \del A_i + A_j \del_i w_1^j ~~~~~~~~~~&&\nonumber\\
- \half w_1^k w_1^l \del_k\del_l a_i - w_1\cdot \del a_j \del_i w_1^j
&\approx&0\label{ab6}
\eeqar
By a suitable choice of the gauge function $\theta$, the first of these equations can be brought to the form
\beq
w_1^j \omega_{ij} + A_i \approx 0
\label{ab7}
\eeq
which has the simple solution
\beq
w_1^j = - \omega^{-1jk} A_k
\label{ab8}
\eeq
Equation (\ref{ab6}) can be simplified, again by removal of a suitable $\theta$-factor, as
\beq
w_2^j \omega_{ij} + w_1^j F_{ij} + \half w_1^k w_1^l \del_k \del_l a_i + \del_i w_1^l w_1\cdot \del a_j \approx 0
\label{ab9}
\eeq
Replacing derivatives of $a$ in terms of $\omega$, we can further simplify this as
\beq
\omega_{ij} w_2^j +F_{ij} w_1^j +\half w_1^k w_1^l \del_k \omega_{li} 
+\half w_1^k \del_i w_1^l \omega_{kl}
\approx 0
\label{ab10}
\eeq
and has the solution
\beq
w_2^j = - \omega^{-1jk} \left[ F_{kl} w_1^l + \half w_1^m w_1^n \del_m \omega_{nk} 
+ \half w_1^m \del_k w_1^n \omega_{mn} \right]
\label{ab11}
\eeq

These results can now be used to calculate the new Hamiltonian as given by
(\ref{ab2}), to the second order in the $A$'s and derivatives. We find
\beqar
H &=& {\cal A} = (V+A_0)\biggr]_{v-w}\nonumber\\
&=& V+A_0 - w_1\cdot \del (V+A_0) +\half w_1^k w_1^l \del_k \del_l (V+A_0)
- w_2 \cdot \del (V+A_0) +\cdots\nonumber\\
&=& V+A_0 + \omega^{-1jk} A_k \del_j (V+A_0) + \half \omega^{-1ik} \omega^{-1jl} A_k A_l \del_i \del_j V \nonumber\\
&&\hskip .3in+
\omega^{-1jk} \del_j V \left[ F_{kl} w_1^l + \half w_1^m w_1^n \del_m \omega_{nk}
+\half w_1^m \del_k w_1^n \omega_{mn} \right] +\cdots\label{ab12}
\eeqar
We shall now define $u^i = \omega^{-1 ij} \del_j V$. This would be the velocity
(or ${\dot v}^i$) if the Hamiltonian were just $V$. Second derivatives like
$\del_i \del_j V$ can now be written in terms of derivatives of $u$, with a compensating term involving
derivatives of $\omega$. We also use the expression for $w_1$ from (\ref{ab8})
to bring the Hamiltonian to the form
\beqar
H \equiv {\cal A} &=& V +A_0 - u\cdot A + \omega^{-1ij} \del_i A_0 A_j - \half \omega^{-1ik} A_k A_j \del_i u^j \nonumber\\
&&-\half u^k w_1^m w_1^n \left[ \del_m \omega_{kn} + \del_k \omega_{nm} 
+\del_n \omega_{mk}\right] 
+ u^k \omega^{-1ln} A_n F_{kl} \nonumber\\
&&+\half w_1^k (u\cdot \del A_k) +\cdots \label{ab13}
\eeqar
Using the fact that $\omega$ is closed, i.e., $ \del_m \omega_{kn} + \del_k \omega_{nm} 
+\del_n \omega_{mk} =0$, and with some further simplification, this gives
\beqar
{\cal A} &=& V +A_0 - u\cdot A + \omega^{-1ij} \del_i A_0 A_j - \half \omega^{-1ij} A_j A_k ~\del_i u^k
+\half \omega^{-1ij} A_j~ u\cdot \del A_i \nonumber\\
&&\hskip .3in- \omega^{-1ij} u^k A_j \del_i A_k
+\cdots\label{ab14}
\eeqar

Since we just used $V+A_0$ without the time-derivative term, this procedure does not fix the terms 
in the Hamiltonian which involves time-derivatives. We can obtain these terms by noting that
$A_0$ is defined only up to a gauge tarnsformation. Thus $A_0$ and $A_0 +\del_0 \Lambda$ should
give ${\cal A}$'s which differ by $\del_0 \Lambda$. This identifies the correction term as
$\half \omega^{-1ij} A_i \del_0 A_j$.
The total result is thus
\beqar
{\cal A} &=& V +A_0 - u\cdot A + \omega^{-1ij} \left[ \del_i A_0 A_j +\half A_i \del_0 A_j\right]\nonumber\\
&&\hskip .3in+\omega^{-1ij} \left[ \half A_j u\cdot \del A_i  - u^k A_j \del_i A_k
-\half A_j A_k \del_i u^k\right]
+\cdots\label{ab15}
\eeqar
This is the basic result for Abelian fields. It agrees with the result in \cite{karabali}
with the identification $\omega = - n \Omega$ of that paper.

The expression (\ref{ab15}) can be related to Chern-Simons action as follows.
The $(2k+1)$-dimensional Chern-Simons action is given by
\beq
S_{CS} = {i^{k+1}\over (2\pi )^{k} k!} \int  A (dA)^k
\label{ab16}
\eeq 
We replace $A$ by $a+A$ and expand this to obtain
\beq
S_{CS} = {i^{k+1}\over (2\pi )^{k} k!} \int  \left[ \omega^k A_0 + {k \over 2}
\omega^{k-1} ~AdA
+ {k (k-1) \over 3!} \omega^{k-2} ~AdA dA +\cdots \right]
\label{ab17}
\eeq
In equations (\ref{ab16}) and (\ref{ab17}), we have used antihermitian
fields, to be in conformity with our later discussions of the nonabelian case.
Thus, in these equations, $\omega $ is given in terms of the
K\"ahler form $\omega_K$ on ${\bf CP}^k$ as $\omega = -i n \omega_K$ and
the fields are given as $-i$ times the hermitian fields.
For the product of the (real) components of $\omega$'s we can use the rules
\beq
\epsilon^{a_1 a_2 \cdots a_{2k} } \omega_{a_1 a_2} \cdots \omega_{a_{2k-1} a_{2k}}
= 2^k k! \sqrt {\det \omega}\nonumber
\eeq
\beqar
\epsilon^{ij a_1 a_2 \cdots a_{2k-2} } \omega_{a_1 a_2} \cdots \omega_{a_{2k-3} a_{2k-2}}
&=&  2^k k!  \sqrt {\det \omega}~\biggl[ -{1\over 2k} \omega^{-1ij}\biggr]\label{ab18}\\
\epsilon^{ijkl a_1 a_2 \cdots a_{2k-4} } \omega_{a_1 a_2} \cdots \omega_{a_{2k-5} a_{2k-4}}
&=&  2^k k! \sqrt {\det \omega} ~\biggl[ {1\over 2^2 k (k-1)}  \bigl( \omega^{-1ij} \omega^{-1kl}
\nonumber\\
&&\hskip .6in+ \omega^{-1ik} \omega^{-1lj} + \omega^{-1il} \omega^{-1jk}\bigr)
\biggr]\nonumber
\eeqar
We also have
\beq
\omega^k = (-in)^k \omega_K^k = (-in)^k (2\pi )^k d\mu ({\bf CP}^k)
\label{ab18a}
\eeq
Simplifying (\ref{ab17}) using these results, and writing everything in terms of real
or hermitian fields, we find
\beqar
S_{CS} &=& {n^k \over k!} \int dt ~d\mu ({\bf CP}^k)~ \biggl[ A_0 + \omega^{-1ij}
\biggl( \half  A_i \del_0 A_j - A_i \del_j A_0 \biggr)\nonumber\\
&&\hskip 1in +\biggl( \half A_i \del_j A_0 \del_k A_l - \onethird  A_i \del_0 A_j \del_k A_l
\biggr) \nonumber\\
&&\hskip 1in \times~ \biggl( \omega^{-1ij} \omega^{-1kl}+ \omega^{-1ik} \omega^{-1lj} + \omega^{-1il} \omega^{-1jk}\biggr)+\cdots\biggr]
\label{ab19}
\eeqar
In this equation, we now replace $A_0$ by $A_0 +V$ and again expand to quadratic terms in the $A$'s keeping in mind that $V$ is taken to be of order zero. This gives, finally,
\beqar
S_{CS}(V+ A_0 , a+A) &=&  { n^k \over k!}  \int dt~d\mu ({\bf CP}^k)~\biggl[ V+ A_0 
- u\cdot A \nonumber\\
&&\hskip .5in+ \omega^{-1ij}\biggl( \half A_i \del_0 A_j - A_i \del_j A_0 \biggr)\nonumber\\
&&\hskip .5in +\omega^{-1ij}  \biggl( u^k A_i \del_j A_k - \half \del_i u^k A_j A_k 
-\half A_i u\cdot \del A_j \biggr) +\cdots\biggr]\nonumber\\
&=&  { n^k \over k!}  \int dt~ d\mu ({\bf CP}^k)~ {\cal A}
\label{ab20}
\eeqar
It may b worth emphasizing that, in this expression, $\omega^{-1} = \omega^{-1}_K/n$.
Equation (\ref{ab20}) gives the relation between the Chern-Simons action and
${\cal A}$ in (\ref{ab15}).
Expressions
(\ref{ab15}) and (\ref{ab20}) are in complete agreement with the expressions
in \cite{karabali}. (In making this comparison, it should be kept in mind that 
$n\Omega$ in \cite{karabali} is the negative of our $\omega$. This is due to the
definition of the total field in \cite{karabali} as $-a +A$, rather than $a+A$.)

Notice also that the phase volume $d\mu_P$ is given by
\beq
d \mu_{P} = ~{n^k \over k!}~ d\mu ({\bf CP}^k)
\label{ab21}
\eeq
with $n^k /k!$ being the number of states in the large $n$ limit. Thus the basic result 
of this section can be expressed as
\beq
S_{CS} = \int dt ~d\mu_{P} ~ {\cal A}
\label{ab22}
\eeq

\section{General matrix formulation}

\subsection{General framework and results}

The previous derivation is expressed entirely in classical language, without considering transformations directly on the Hilbert space. Secondly, it is not clear how the arguments can be extended to nonabelian fields. While it is possible to write symplectic forms including
nonabelian degrees of freedom, as one does for the Wong equations of motion,
the kind of coordinate changes we need to make and their implementation in terms of matrix
representation of the nonabelian algebra are not obvious. In any case, we want to take the point of view that the Hilbert space is the fundamental entity, with the classical description being just a large $n$ simplification. This is also the proper starting point if we think of 
the Hilbert space as defining a fuzzy version of the space.
In such cases, it is important to have a matrix version of the calculations of the previous section; this section will address this question. 

The action we start with is a matrix action and therefore, there is, initially, no notion of space or spatial geometry.
The Hilbert space ${\cal H}_{\mathcal N}$ on which the matrices act as linear transformations can be taken as
arising from quantization of ${\bf CP}^k$ for one, or different values of $k$, or from some other phase space. Even for the same geometry and topology for the phase space, the choice of symplectic form is not unique.
For example, one could use a specific form $\omega$ or some perturbation of it which 
does not change the phase volume so that the total number of states upon quantization does  not change.
For each such possibility there is a corresponding
large ${\mathcal N}$ limit. Thus the same matrix action can be
expanded out in many ways for large ${\mathcal N}$.
We shall work this out by expanding around some specific choice of background, which is the analogue of the $a$ of the previous section, with extra gauge potentials $A$.
The final result will not be sensitive to the choice of background, so that it can be used for
a number of different backgrounds.

The matrix version of the action which we consider is given by
\beq
S = i \int dt~ \Tr \Bigl( \rho_0 U^\dagger D_0 U \Bigr)\label{GM1}
\eeq
where $\rho_0$ denotes the density matrix of the system initially and $A_0$ is related to
the Hamiltonian. From now on we will use properly antihermitian $A_0$'s in line with the spatial components; thus $D_0 = \del_0 +A_0$ and the Hamiltonian is given by $H
=-iA_0$. 
The action (\ref{GM1}) leads to the correct evolution equation for the
density matrix, namely, 
\beq
i D_0 \rho = i {\del \rho \over \del t} +i [A_0 , \rho ] = i {\del \rho \over \del t}
- [H,\rho ]= 0
\label{GM2}
\eeq
where $\rho = U^\dagger \rho_0 U$ is the density matrix at
time $t$. 
Eventually we can take $A_0 \rightarrow A_0 +i V$ to include a background potential.

The unitary transformation $U$ encodes the 
fluctuations of the chosen density matrix, or the edge states in the quantum Hall
way of thinking about this problem. Equivalently, it gives the boundary effects for dynamics in a subspace of a fuzzy space.
The bulk dynamics is not sensitive to $U$, and can be extracted by taking $\rho = {\bf 1}$. Effectively, we are then seeking
the simplification of $S = i \int dt~\Tr D_0$ in the limit of large matrices.
This action is the one-dimensional Chern-Simons action for the matrix theory.

In the following, the specific background we choose corresponds to ${\bf CP}^k$.
The gauge fields $A_0$, $A_i$ will be expanded around this background; thus
the fields $A_0$, $A_i$ are actually functions on fuzzy ${\bf CP}^k$.
To carry out the expansion, we write $A_0$, $A_i$ in terms of $(N\times N)$-blocks.
In other words, we can take ${\cal H}_{\mathcal N} = {\cal H}_{N} \otimes {\cal H}_M$ so that the matrix elements of $A_i$, $i =0, 1, 2, ...$, may be written as
$A_{i AB} = \la A \vert A_i \vert B\ra = \la \alpha ~a \vert A_i \vert \beta~ b\ra$,
$\alpha ,\beta = 1, 2, \cdots, N$, $a, b = 1, 2, \cdots, M$.
${\cal H}_{N}$ will carry an irreducible representation of $SU(k+1)$,
specifically the symmetric rank $n$ representation.

The chosen background has an Abelian background field which corresponds to
$\omega = -in \omega_K$. A general gauge field is introduced
by the prescription $D_a = -iT_a +A_a$. This will involve $k^2 +2k$ spatial
components for the gauge potential, which are obviously too many for
${\bf CP}^k$. Thus there are restrictions on $D_a$ which ensure that there are
only $2k$ spatial components for the potentials. These conditions may be taken as
the gauged version of the conditions (\ref{res22}), 
\beqar
D_a D_a &=& - C_n \nonumber\\
d_{abc} D_b D_c&=& -i \alpha_n  D_a\label{GM3}
\eeqar
In other words, even after gauging, the derivatives obey the same embedding
conditions (\ref{res22}) as before gauging \cite{KNP}. (In the limit of a continuous manifold, there is some
redundancy in these conditions. While they are
sufficient for our purpose, whether they are necessary and sufficient
in the noncommutative case is not quite a settled issue.)

With this set-up, 
the simplification of the matrix action can now be carried out. We will consider the variation
of the matrix action $i\Tr D_0$ under a change of the background fields.
$D_0$ is a matrix function, and so, it can be expressed in terms of a basis
made up of powers of $T_a$. Since  $T_a = D_a - A_a$, it is then possible to expand
$D_0$ in terms of the $D_a$'s which give the same basis on a background with
additional gauge fields $A_a$.
In a large $n$-expansion, we can also replace the matrices by their symbols.
The resulting action is then given by
\beq
i\int dt~ \Tr (D_0) \approx S_{*CS}~+~\cdots
\label{GM4}
\eeq
where $S_{CS}$ is the Chern-Simons $(2k+1)$-form on ${\bf CP}^k\times {\bf R}$.
$S_{*CS}$ is the star-version of the same action, i.e., $S_{CS}$ with star products connecting the various fields in it. The gauge potentials in the Chern-Simons action are given by
$a +A$, where $a$ is the background value corresponding to the symplectic form
and $A$ is the additional potential or gauge field fluctuation.

The integer $n$ may be written in terms of the scale parameter $R$, which is the
analog off the radius of the manifold, as $n =2 BR^2$, $B$ being the field strength for the 
symplectic form.
If we take the large $R$ limit, and further take the gradients of the fields to be small compared to their values, so that $D^2 F \ll F F $, for example, then we can simplify the result (\ref{GM4})
as
\beq
i\int dt~ \Tr (D_0) \approx S_{CS}~+~\cdots
\label{GM5}
\eeq
In other words, in the spirit of a gradient expansion, and at large $R$, the
higher terms in the star product are negligible. 

Equations (\ref{GM4}) and (\ref{GM5})
are the main results we will show over next few subsections.
However,
the general analysis is algebraically rather involved, with
many calculational steps. Therefore, before
continuing with the general matrix formulation,
it is useful to work out the case of a two-sphere. This will exemplify the 
calculational steps required for the general analysis.

\subsection{The effective action on the two-sphere}

In the case of the two-sphere, we need rank $n$ $SU(2)$ representations, so that 
$N =n+1$.
The gauge fields are written in terms of $(n+1)\times (n+1)$-blocks as
$A_{i AB} = \la A \vert A_i \vert B\ra = \la \alpha ~a \vert A_i \vert \beta~ b\ra$,
$\alpha ,\beta = 1, 2, \cdots, n+1$, $a, b = 1, 2, \cdots, M$.
The large $n$ analysis is facilitated by the use of the symbol which we define as
\beq
(A_i)_{ab} = \sum_{\alpha , \beta} {\cal D}^{(n)*}_{\alpha , n} (g)~ \la \alpha ~a \vert A_i \vert \beta~ b\ra ~{\cal D}^{(n)}_{\beta ,n}(g)
\label{sph1}
\eeq
where ${\cal D}$'s are the Wigner ${\cal D}$-functions for $SU(2)$.
$(A_i )_{ab}$ is thus a matrix-valued function 
on $S^2$; it takes values in the Lie algebra of $U(M)$ and, in the large $n$ limit, it can be identified as the gauge potential on $S^2\times {\bf R}$, where
${\bf R}$ is the temporal dimension.

A parenthetical remark is in order at this point. In the case of higher dimensional group coset manifolds of the form $G/H$, where $H$ is nonabelian, such as 
${\bf CP}^k$, it is possible to choose a nonabelian background field \cite{KN}.
The relevant wave functions are of the form ${\cal D}^{(J)}_{A a}(g)$, where $J$ denotes a
representation of $G$, and we have
a nontrivial representation of $H$ for the right translations of $g$, corresponding to the index $a$.
The symbol can then be defined as
\beq
(A_i)_{ab} = \sum_{A, B} {\cal D}^{(J)*}_{A , a} (g)~ \la A \vert A_i \vert B \ra ~{\cal D}^{(J)}_{Bb}(g)
\label{sph1a}
\eeq
This was the definition used in \cite{KN} and \cite{karabali}. It is a little bit different
from (\ref{sph1}). We will use (\ref{sph1}). In the simplification of the action,
this will lead to traces over the remaining matrix labels
such as $a, b$ in (\ref{sph1}). At this stage, if $G\in U(M)$, we can rewrite the $M\times M$-matrix 
products and traces by introducing 
additional ${\cal D}$'s and
using $\delta_{ab}  = {\cal D}^{K*} _{ac} {\cal D}^K_{bc}$, where ${\cal D}^K$'s are 
suitable representations
of $G$. (More than one irreducible representation may be needed) . We can then
reduce the product ${\cal D}^{(n)}_{\beta ,n}
{\cal D}^K_{bc}$ to get wave functions like ${\cal D}^{(J)}_{Bc}$. Results using (\ref{sph1a})
can thus be recovered from results obtained using (\ref{sph1}).
This argument also shows that at least a part of the remaining gauge group
can be identified with gauge fields for
the isometries, namely the group $G$, of the space. The indices $a, b$ in
(\ref{sph1}) play the role of the tangent space indices.

Returning now to the main line of reasoning, let $T_\mu$ be $(n+1)\times (n+1)$-matrices which are the generators of the $SU(2)$ Lie algebra in the $j = \half n$ representation. The covariant derivative of a function $f$
can be written as
$D_\mu f = [ -iT_\mu , f] + A_\mu f$. Evidently the covariant derivatives obey
\beq
[D_\mu , D_\nu ]=  f_{\mu\nu\alpha} D_\alpha ~+~ F_{\mu\nu}\label{sph2}
\eeq
where
\beq
F_{\mu\nu} = [ -iT_\mu , A_\nu ] - [ -iT_\nu , A_\mu ]  + [A_\mu , A_\nu ] - f_{\mu\nu\alpha}
A_\alpha
\label{sph3}
\eeq

For the matrices $T_\mu$ we have the condition $T_\mu T_\mu = C_n =
{1\over 4} n (n+2)$, where
$C_n$ is the quadratic Casimir for the 
representation with $j =\half n$. We can define $X_\mu = T_\mu /\sqrt{C_n}$ as the 
coordinates of the fuzzy sphere; they correspond to the fuzzy version
of the coordinates
one has by embedding the two-sphere in ${\bf R}^3$, since $X_\mu X_\mu =1$. 
This is the $k=1$ case of the equations (\ref{res21}).
At the level of the $D$'s, notice that
there are three $D$'s and three $A$'s, whereas we should have
only two independent ones for the two-sphere. We must have a constraint
which is the analogue of the condition $T^2 = C_n$. The required
constraint may be taken as $D_\mu D_\mu = - C_n$. In other words, even after the gauging
$-iT_\mu \rightarrow -iT_\mu +A_\mu = D_\mu$, we must have the sum of squares
equal to the quadratic Casimir value.

In the following, we shall denote the commutator $[D_\mu , D_\nu ]$ by 
$\Omega_{\mu\nu}$. Explicitly, $\Omega_{\mu\nu} = f_{\mu\nu\alpha} D_\alpha
+F_{\mu\nu}$. 
$\Omega_{\mu\nu}$ is not invertible in general, but on the restricted subspace with
fixed value of $D^2$, it has an inverse; equivalently, we can find
a matrix $N_{\mu\nu}$, at least as a series in $F$'s and their derivatives, such that
\beq
N_{\mu\nu}  \Omega_{\nu\alpha} = \delta_{\mu\alpha} + {1\over C_n}D_\alpha D_\mu 
\label{sph4}
\eeq
Notice that, at least in the large $n$ limit, the right hand side
of (\ref{sph4}) is a projection operator to directions tangential to the
sphere.
If we keep just the first term in $\Omega_{\mu\nu}$, namely
$\omega_{\mu\nu} \equiv f_{\mu\nu\alpha} D_\alpha$, we find
\beq
N_{\mu\nu} \approx N_{0\mu\nu} = {1\over C_n} f_{\mu\nu\alpha } D_\alpha
\label{sph5}
\eeq
In writing $\Omega_{\mu\nu} = f_{\mu\nu\alpha} D_\alpha + F_{\mu\nu} $, the second term,
$F_{\mu\nu}$ is of lower order in powers of $n$ compared to the first one, since
$D$ is of order $n$. (Recall that $D^2 = -C_n$ is of order $n^2$.)
Thus $N_{0\mu\nu}$ is the leading term we need at large $n$, the corrections to
(\ref{sph5}) are subdominant. They will be needed for higher dimensional spaces,
as we shall see later.

Using equation (\ref{sph4}), we find
\beq
\delta D_\alpha =  {1\over 2}\bigl( \delta D_\mu N_{\mu\nu} [D_\nu , D_\alpha ] +
[D_\alpha , D_\nu ] N_{\nu\mu} \delta D_\mu \bigr)- {1\over 2 C_n}
\bigl( \delta D_\mu D_\alpha D_\mu + D_\mu D_\alpha \delta D_\mu \bigr)
\label{sph6}
\eeq
The second set of terms on the right hand side may be simplified as
\beqar
- {1\over 2 C_n}
\bigl( \delta D_\mu D_\alpha D_\mu + D_\mu D_\alpha \delta D_\mu \bigr)
&=& {1\over 2C_n}  \bigl( \delta D_\mu \delta_{\mu\nu} [D_\nu , D_\alpha ] +
[D_\alpha , D_\nu ] \delta_{\nu\mu} \delta D_\mu \bigr)\nonumber\\
&&\hskip .2in
-{1\over 4C_n}[ \delta D \cdot D - D-\delta D , D_\alpha ] \nonumber\\
&&\hskip .2in
-{1\over 4 C_n} \bigl( \delta D^2 ~ D_\alpha + D_\alpha ~\delta D^2 \bigr)
\label{sph7}
\eeqar
Since $D^2$ is fixed, the last term is actually zero. Using this equation, we find
\beqar
\delta D_\alpha &=&  {1\over 2}\bigl( \xi_\nu [D_\nu , D_\alpha ] +
[D_\alpha , D_\nu ]{\tilde\xi}_\nu \bigr)
-{1\over 4C_n}[ \delta D \cdot D - D\cdot\delta D , D_\alpha ]\nonumber\\
\xi_\nu &=&\delta D_\mu M_{\mu\nu} , \hskip .4in 
{\tilde \xi}_{\nu}  = M_{\nu\mu}\delta D_\mu
\label{sph8}\\
M_{\mu\nu} &=&N_{\mu\nu}~+~  {\delta_{\mu\nu}\over C_n} \nonumber
\eeqar
The second term on the right hand side gives a unitary transformation of
$D_\alpha$.

Let $K$ be a matrix acting on the Hilbert space ${\cal H}$. 
We may write
$K$ in an expansion in $D$'s as a sum of terms of the form
\beq
K = K^{\mu_1 \mu_2 \cdots \mu_s} ~D_{\mu_1} D_{\mu_2} \cdots D_{\mu_s}
\label{sph9}
\eeq
(Our results extend by linearity to sums of such terms, so it is sufficient to consider one such term,
for a fixed value of $s$.)

In the large $n$ limit, the $D$'s typically become the coordinates for the
space ${\cal M}$, in an embedding of ${\cal M}$ in ${\bf R}^d$ of suitable dimension $d$,
as indicated after equation (\ref{res20}).
The question of interest is thus how  we can express $K$ in an expansion around
a perturbed version of the $D$'s, namely, $D'_\mu $ where $A'_\mu = A_\mu +\delta A_\mu $. Clearly this can be achieved
by writing $D_\mu = D'_\mu - \delta A_\mu = D'_\mu - \delta D_\mu$.
The change in $K$ is thus given by $D_\mu \rightarrow D_\mu -\delta D_\mu$.
Using equation (\ref{sph8}), we may write
\beq
\delta K = \half \bigl( \delta_1 K + \delta _2 K\bigr)
\label{sph10}
\eeq
where
\beqar
\delta_1 K \!\!\!&=&\!\!\! -K^{\mu_1 \mu_2 \cdots \mu_s} \Bigl[ \xi^\alpha [D_\alpha , D_{\mu_1}] D_{\mu_2}
\cdots D_{\mu_s} + D_{\mu_1} \xi^\alpha [D_\alpha , D_{\mu_2}] D_{\mu_3}\cdots D_{\mu_s}
+\cdots \Bigr]\nonumber\\
\delta_2 K \!\!\!&=&\!\!\! -K^{\mu_1 \mu_2 \cdots \mu_s} \Bigl[ [D_{\mu_1}, D_\alpha ] {\tilde
\xi}^\alpha D_{\mu_2}
\cdots D_{\mu_s} + D_{\mu_1}  [D_{\mu_2}, D_\alpha ] {\tilde
\xi}^\alpha D_{\mu_3}\cdots D_{\mu_s}
+\cdots \Bigr]\nonumber\\
\label{sph11}
\eeqar
We have not written the unitary transformation by $(\delta D \cdot D - D \cdot \delta D)$
which will lead to a unitary transformation of $K$. Eventually we take a trace to get the action, so this will not affect our considerations for the bulk terms. If we were to keep the boundary terms, such a term would be important.
If all $\xi$'s are moved to the left end of the expression for $\delta_1K$, we will get
$\xi^\alpha [ D_\alpha , K]$. This will be the leading term at large $n$. In moving the
$\xi$'s to the left, we will get extra commutator terms, which will be related to Poisson brackets at large $n$, and are subdominant. The commutator terms will be discussed later in this section, but for now, we focus on the dominant terms at large $n$ and write,
\beq
\delta K \approx  -{1\over 2} \bigl( \xi^\alpha [D_\alpha , K] + [K, D_\alpha ] {\tilde \xi}^\alpha
\bigr)
\label{sph12}
\eeq

We are now in a position to apply this line of reasoning to the computation of the effective action (\ref{GM1}). As mentioned before, the bulk dynamics is not sensitive to $U$, and 
for obtaining the bulk action we can set $\rho = {\bf 1}$. 
Thus we need the simplification of $\Tr D_0$ in the limit of large matrices.
We start with the expansion of $D_0$ around a background with
potential $A_\mu +\delta A_\mu$.
This can be worked out by taking $K=D_0$ in the above equations.
The first correction due to change of $A_\mu$ is found as
\beqar
\delta D_0 &=& -{1\over 2} \Bigl[ \delta D_\mu M_{\mu\nu} [D_\nu, D_0]
+ [D_0, D_\nu ] M_{\nu\mu} \delta D_{\mu} \Bigr]\nonumber\\
&=& -{1\over 2} \Bigl( \delta A_\mu M_{\mu\nu}~ F_{\nu 0}
- F_{\nu 0} ~M_{\nu\mu} \delta A_\mu \Bigr)
\label{sph15}
\eeqar

In using the expression for $M_{\mu\nu}$ from (\ref{sph8}), the contribution
of $\delta_{\mu\nu}/C_n$ is zero by cyclicity of trace. Also keeping just the
$N_{0\mu\nu}$-term in $N_{\mu\nu}$ as explained after (\ref{sph5}),
we find
\beqar
\delta S &\approx& -i {(n+1)\over 2} \int dt d\mu ({\bf CP}^1)~\tr \bigl(
\delta A_\mu N_{0\mu\nu} F_{\nu 0} - F_{\nu0} N_{o\nu\mu} \delta A_\mu \bigr)\nonumber\\
&\approx& - {n(n+1) \over 2 C_n} \int dt d\mu ({\bf CP}^1)~f_{\mu\nu\alpha}
S_{\alpha 3}~ \tr \bigl(\delta A_\mu F_{\nu 0}\bigr)
\label{sph16}
\eeqar
We have used equation (\ref{res17}) to go to the integral over phase space.
We have also used the fact that
\beq
N_{0\mu\nu} = {1\over C_n}  f_{\mu\nu\alpha} D_\alpha \approx -{i n\over 2 C_n}
f_{\mu\nu\alpha} S_{\alpha 3}
\label{sph17}
\eeq 
Further, we have replaced the matrices by symbols, so that $\delta A_\mu$
and $F_{\nu0}$ in this equation are $M\times M$-matrices.

Further simplification is as follows. Notice that $f_{\mu\nu\alpha} S_{\alpha 3}
= f_{ab3} S_{\mu a} S_{\nu b}$ from the definition of $S_{\mu\alpha}$.
From (\ref{res20}), we see that $S_{\nu b} F_{\nu 0}
= F_{b0} = [iR_b , A_0] - \del_0 A_b +\cdots$. $F_{b0}$ is in a basis
where the derivatives are the $R_i$'s. We can convert these to  the standard coordinate basis
by writing
$S_{\nu b} F_{\nu 0} = (E^{-1})^i_b F_{i0}$ where $E^a_i$ is the frame field for the metric.
This yields
\beqar
\delta S &\approx&   - {n(n+1) \over 2 C_n} \int dt d\mu ({\bf CP}^1)~
\epsilon_{ab} (E^{-1})^i_a (E^{-1})^j_b ~\tr \bigl( \delta A_i F_{j0}\bigr)\nonumber\\
&\approx& - {1\over 2\pi} \int dt d\mu (S^2)~ (\det E)^{-1} \epsilon^{ij}~
\tr \bigl( \delta A_i F_{j0}\bigr)
= - {1\over 2\pi} \int dt ~\tr \bigl( \delta A  F \bigr)
\label{sph18}
\eeqar
We had normalized the volume $d\mu ({\bf CP}^1)$ to unity; so, in
the second step, we changed to the standard normalization of $4\pi$.
(We have also used $C_n \approx n^2 /4$ and, in the last step, changed to
differential forms for the fields.)

The variation of the $(2k+1)$-dimensional Chern-Simons term is given by
\beq
\delta S_{CS} =  {i^{k+1}\over (2\pi )^k k!} \int \tr (\delta A F^k)
\label{sph19}
\eeq
For $k=1$, we find $\delta S = - (1/2\pi ) \int \tr (\delta A F)$, so that, comparing with
(\ref{sph18}) we conclude
\beq
S \equiv i \int dt ~\Tr ( D_0 ) \approx S_{CS} 
\label{sph20}
\eeq

A number of remarks are in order at this point.
At finite values of $n$, the Hilbert space ${\cal H}_N$ defines a fuzzy version of
the phase space, specifically, in this case, the two-sphere. Thus the action
$\int dt~ \Tr (D_0)$ is a matrix model action, or equivalently a fuzzy space action.
What we have shown is that the large $n$ limit is given by the
continuum Chern-Simons theory on $S^2 \times {\bf R}$.
(Notice that, as should be expected, the result is independent of the
metric on $S^2 \times {\bf R}$, only the topology is relevant in this action.)
Chern-Simons theories have been formulated and analyzed on many noncommutative spaces
such as the fuzzy sphere \cite{poly2} and the noncommutative plane \cite{poly}. 
For example, on the plane, the action is given by
\beq
S = -{1\over 2\pi} \left[ ~{\theta\over 3}~ \Tr (D_\mu D_\nu D_\alpha ) \epsilon^{\mu\nu\alpha}
-i \Tr (D_0) \right] 
 \label{sph21}
\eeq 
These actions are different from $\int dt \Tr (D_0)$ at the matrix level,
and explicitly involve the spatial components of the gauge field.
However, their commutative limits lead to the 
standard commutative Chern-Simons action.
What we have shown is that the action $\int dt~ \Tr (D_0)$, expanded in a suitable background, also leads to the same commutative limit.
This is not surprising if one views the finite $n$ case as a regularization of the continuum theory. It is, of course, possible to have many theories which 
are different at the regulated level but which coincide when the regulator is 
eventually removed.
One can study the different noncommutative Chern-Simons theories as genuinely different theories at the matrix level. But if we only ask for a matrix theory whose commutative limit
gives the Chern-Simons theory, then the action (\ref{GM1}) is a good choice.
In this sense, it defines a fuzzy version of the Chern-Simons theory.

In generalizing this construction, it is clear that there are three basic ingredients
that we must take account of. First of all, we need to show that a suitable
construction of the matrix $N_{\mu\nu}$ can be made for the full $\Omega_{\mu\nu}$, at least as an expansion in powers of $1/n$. Secondly, the extra terms in equations
(\ref{sph11}) which arise from reordering must be calculated. And finally,
the definition of the symbol
is altered when we make the shift $A_\mu\rightarrow A_\mu +\delta A_\mu$, and this must be incorporated at the step of replacing matrices by their symbols. In the next three subsections, we will address these issues.

\subsection{Construction of $N_{ac}$}

We start with equation (\ref{sph2}) again, $[D_a, D_b] = f_{abc} D_c
+F_{ab} \equiv \Omega_{ab}$. First, we want to obtain $N_{ac}$ in a form suitable for large
$n$ analysis. For this purpose, we define a gauge covariant quantity $R_{ab}$ by the equation
\beq
f_{ack} D_k~ f_{cbl}  D_l = D_b D_a + {C_n \over k} \delta_{ab}
+ i {2n +k +1 \over 2} d_{abc} D_c 
- {k-1 \over 4} f_{abc} D_c ~+~R_{ab}
\label{N1}
\eeq
The key observation is that $R_{ab}$, which is naively of order $n^2$, 
is actually of lower order in powers of $n$. This follows from the fact that in
$D_a = -iT_a +A_a$, $T_a$ is of order $n$, but $A_a$ is of order $1$. If $A_a =0$,
we just have the $T$-matrices and, in the symmetric rank $n$ representation, they obey the
identity
\beq
- f_{ack} T_k~ f_{cbl}  T_l = - T_b T_a + {C_n \over k} \delta_{ab}
+  {2n +k +1 \over 2} d_{abc} T_c 
+i {k-1 \over 4} f_{abc} T_c 
\label{N2}
\eeq
Thus $R_{ab}$ vanishes to leading order in $n$. It does not vanish at the next order,
but its contribution to the action will be down by two powers of $n$ compared to the leading term.) 

There are other, equivalent, ways of writing $R_{ab}$. For example, the identity
\beq
f_{cak}~f_{cbl}
= {2\over k+1} \left( \delta_{ab}\delta_{kl}
-\delta_{al}\delta_{kb}\right) + 4 \left(d_{abm} d_{klm} - d_{alm} d_{kbm}\right)
\label{N3}
\eeq
leads to
\beqar
R_{ab} &=& (k-1) {C_n\over k+1} \left( {\delta_{ab}\over k} + {D_b D_a \over C_n}
\right) +i {(k-3) (2n+k+1) \over 2 (k+1)} d_{abc} D_c\nonumber\\
&&\hskip 1.3in
+ 4~ d_{alm} d_{bkm} D_k D_l +{k-1\over  4} f_{abc} D_c
\label{N4}
\eeqar
Going back to (\ref{N1}), we rewrite it as
\beq
\biggl[ f_{ack} D_k + {1\over 4} (k-1) \delta_{ac}\biggr]~ f_{cbl}  D_l = D_b D_a + {C_n \over k} \delta_{ab}
+ i {2n +k +1 \over 2} d_{abc} D_c 
~+~R_{ab}
\label{N5}
\eeq
We now define a quantity
\beq
{\bf Y}_{ab} = d_{abc} D_c +i {(k-1) (2n +k +1)\over 8 (k+1)} ~\delta_{ab}
= d_{abc} D_c +i { \alpha_n\over 2} \delta_{ab}
\label{N6}
\eeq
Equation (\ref{N5}) now becomes
\beq
\biggl[ f_{ack} D_k + {1\over 4} (k-1) \delta_{ac}\biggr]~ f_{cbl}  D_l = D_b D_a + 
B_n \delta_{ab}
+ i {2n +k +1 \over 2} {\bf Y}_{ab}
~+~R_{ab}
\label{N7}
\eeq
where
\beq
B_n = {n (n+k+1) \over 4} + {k^2 -1 \over 16}
\label{N8}
\eeq
Notice that $B_n = C_n$ for $SU(2)$ for which $k=1$. We now define
\beq
N_{0ac} = {1\over B_n} \biggl[  f_{ack} D_k + {1\over 4} (k-1) \delta_{ac}\biggr]
\label{N9}
\eeq
Equation (\ref{N7}) can now be written as
\beqar
\bigl( N_0 \omega )_{ab} 
&=&\delta_{ab} + {\mathbb X}_{ab} + i {\mathbb Y}_{ab} ~+~ {\mathbb R}_{ab}
\nonumber\\
{\mathbb X}_{ab} = {D_b D_a \over B_n}, \hskip .3in {\mathbb Y}_{ab}
&=& {1\over B_n}  \left( n +{\half} (k+1)\right) {\bf Y}_{ab}, \hskip .3in
{\mathbb R}_{ab} = {R_{ab}\over B_n}
\label{N10}
\eeqar
${\mathbb X}$ and ${\mathbb Y}$ are of order $1$; while ${\mathbb R}$ is naively of the same order, it is actually of lower order, as argued before.
For the full $\Omega_{ab} = f_{abc}D_c  +F_{ab}$, we can now define
$N_{ab}$ by the equation
\beq
(N \Omega )_{ab}  = \delta_{ab} + {\mathbb X}_{ab} + i {\mathbb Y}_{ab} 
\label{N11}
\eeq
$N_{ab}$ can be obtained as a series by writing $N = N_0 + N_1 + N_2+  \cdots$, and matching terms of the same order in powers of $n$ in (\ref{N11}).
The first few terms are given by
\beqar
N_{ab} &=& N_{0ab} - ({\mathbb R}_{ab} N_0 )_{ab}  - (N_0 F N_0 )_{ab} 
+ ( {\mathbb R}({\mathbb X} + i {\mathbb Y} ) N_0 )_{ab}
+( N_0 F({\mathbb X} + i {\mathbb Y} ) N_0 )_{ab}\nonumber\\
&& \hskip 1in + (N_0 F N_0 F N_0)_{ab} +\cdots
\label{N12}
\eeqar
In this series, we encounter terms like ${\mathbb R} N_0$ and ${\mathbb R} {\mathbb X}  N_0$ which are, again, seemingly of the same order, since ${\mathbb X}$ is of order $1$.
However, the product ${\mathbb X}N_0$ reduces the power of $n$ by one, as seen from the following.
\beqar
({\mathbb X} N_0 )_{ab} &=& {D_c D_a \over B_n}\biggl[  {f_{cbk}D_k \over B_n}
+ {\cal O}(1/n^2)\biggr]\nonumber\\
&=& {1\over B_n^2} ~\left(D_a D_c + f_{cam} D_m +F_{ca} \right)~f_{cbk} D_k + {\cal O}(1/n^2)\nonumber\\
&=&  -{1\over 2B_n^2} f_{cbk} D_a [D_c , D_k]  + f_{cam}f_{cbk} {D_m D_k \over B_n^2}
+f_{cbk} {F_{ca}  D_k \over B_n^2} +\cdots\label{N13}
\eeqar
Each term on the right hand side is seen to be of order $1/n^2$, one order less than
what is expected from the order of ${\mathbb X}$ and $N_0$.
In a similar fashion, using the identity
\beq
\biggl[ d_{abc} (-iT_c) +i {(k-1) (2n +k +1)\over 8 (k+1)} ~\delta_{ab} \biggr]
~f_{cbl} (-iT_l) = - {k+1 \over 4} d_{abk} (-iT_k)
\label{N14}
\eeq
we can easily check that ${\mathbb Y}_{ac} N_{0cb}= {\cal O}(1/n^2)$, even though
the left hand side is naively of order $1/n$. In conclusion, the terms
$ {\mathbb R}({\mathbb X} + i {\mathbb Y} ) N_0 $, 
$ N_0 F({\mathbb X} + i {\mathbb Y} ) N_0$ are indeed smaller by a power of $n$, at large $n$, compared to
the corresponding terms $({\mathbb R}_{ab} N_0 )_{ab}$ and 
$(N_0 F N_0 )_{ab} $. Thus the expansion (\ref{N12}) is appropriate at large $n$.

What we have shown is that we can construct a matrix $N$ which obeys equation (\ref{N11});
it is obtained as a series (\ref{N12}), the subsequent terms in the series being smaller
by powers of $n$, as $n$ becomes large.

\subsection{The variation of the action at large $n$}

\noindent{$\underline{Re-expressing ~ the ~variation ~of ~ D}$}

We shall now express the variation of $D_a$ in a form suitable for application to functions.
Equation (\ref{N11}) tells us that 
\beqar
N_{ac} \Omega_{cb}  &=& \delta_{ab} + {\mathbb X}_{ab} + i {\mathbb Y}_{ab} \nonumber\\
\Omega_{bc} N_{ca} &=&  \delta_{ba} + {\mathbb X}_{ba} + i {\mathbb Y}_{ba} 
\label{mat1}
\eeqar
The second equation is the hermitian conjugate of the first. Multiplying the first by $\delta D_a$ on the left and the second by $\delta D_a$ on the right and adding, we get
\beq
{1\over 2} \left( \delta D_a N_{ac} \Omega_{cb} + 
\Omega_{bc} N_{ca} \delta D_a \right) 
= \delta D_b + { \delta D_a D_b D_a + D_a D_b \delta D_a \over 2 B_n }
\label{mat2}
\eeq
In this simplification, we have used the fact that
\beqar
\delta D_a {\mathbb Y}_{ab} + {\mathbb Y}_{ba} \delta D_a
&=& {2 n +k +1\over 2 B_n} \biggl[ d_{abc} (\delta D_a D_c + D_c \delta D_a )
+ i \alpha_n \delta D_b \biggr]\nonumber\\
&=&{2 n +k +1\over 2 B_n}\delta  \biggl[  (d_{abc} D_a D_c )
+ i \alpha_n  D_b \biggr]\nonumber\\
&=&0\label{mat3}
\eeqar 
by virtue of equation (\ref{GM3}).
Equation (\ref{mat2}) is of the same form as (\ref{sph6}) and further simplification is done
using (\ref{sph7}) as before to arrive at
\beqar
\delta D_b &=&  {1\over 2}\bigl( \xi_c [D_c , D_b ] +
[D_b , D_c ]{\tilde\xi}_c \bigr)
-{1\over 4B_n}[ \delta D \cdot D - D\cdot\delta D , D_b ]\nonumber\\
\xi_c &=&\delta D_a M_{ac} , \hskip .4in 
{\tilde \xi}_{c}  = M_{ca}\delta D_a
\label{mat4}\\
M_{ab} &=&  N_{ab}~+~ {\delta_{ab}\over B_n}\nonumber
\eeqar
We have used both the embedding conditions (\ref{GM3}) in arriving at this equation.

\noindent{$\underline{Variation ~of ~a ~matrix ~function ~under ~change~ of ~gauge~ fields}$}

As in the case of $SU(2)$, we can now consider a matrix function
$K$ acting on the Hilbert space ${\cal H}$. $K$ can be taken as the sum of terms of the form $K = K^{a_1 a_2 \cdots a_s} D_{a_1} D_{a_2} \cdots D_{a_s}$, where, we can also take, without loss of generality the coefficients $K^{a_1 a_2 \cdots a_s}$ to be symmetric in all indices. (Any antisymmetric pair may be reduced to a single $D$ and $F$; $F$
itself may be re-expanded in terms of $D$'s, to bring it to this form.)
We may thus write $K$ as
\beq
K = \int d\mu ~ e^{\bz \cdot D } ~K(z) \label{mat5}
\eeq
where
\beq
d\mu = \prod_a {dz_a d\bz_a \over \pi}~e^{-\bz \cdot z}, \hskip .5in
K(z) = K^{a_1 a_2 \cdots a_s} z_{a_1} z_{a_2}\cdots z_{a_s}
\label{mat6}
\eeq
For the variation of $K$ under $D\rightarrow D -\delta D$ we get
\beq
\delta K = - \int d\mu ~ \int_0^1 d\alpha ~ e^{\alpha \bz \cdot D}~ \bz \cdot \delta D ~e^{(1-\alpha )\bz \cdot D} ~ K(z)
\label{mat7}
\eeq
Upon taking traces, this leads to
\beqar
\Tr~ \delta K &=& -\Tr~ (\delta D_a K^a )\nonumber\\
K^a&=& \int d\mu ~\bz^a ~e^{\bz \cdot D} ~K(z) = \int d\mu ~e^{\bz \cdot D}~ {\del K \over \del z_a}\label{mat8}
\eeqar
Substituting from (\ref{mat4}), we get the result
\beqar
\Tr~(\delta K) &=&-{1\over 2} \Tr \biggl[ \delta D_a N_{ab} (\Omega_{bc} K^c)
- \delta D_a (K^c \Omega_{bc}) N_{ba} \nonumber\\
&&\hskip 1.2in +
{1\over B_n} \bigl[\delta D_a (\Omega_{ab} K^b )
-\delta D_a (K^b \Omega_{ab}) 
\bigr] \biggr]
\label{mat8a}
\eeqar
Consider now the commutator $[D_a ,K]$; this can be expanded as
\beqar
[D_a ,K] &=& \int d\mu d\alpha~ e^{\alpha \bz \cdot D} ~[D_a, \bz \cdot D]~
e^{-\alpha \bz \cdot D} ~ e^{\bz \cdot D} ~K(z)\nonumber\\
&=& [D_a , D_b] K^b ~-~ {1\over 2} [[D_a, D_b],D_c]~K^{bc} ~+\cdots
\label{mat9}\\
K^{bc}&=&  \int d\mu~ \bz^b \bz ^c ~e^{\bz \cdot D} ~K(z)\nonumber
\eeqar
Since $D\sim n$ and the commutator reduces the power of $n$ by one, the first term in the
series (\ref{mat9}) is of order $K$, the second of order $K/n$, etc. From the definition of
the coefficients, $[D_a, K^a] =0, ~[D_b, K^{bc} ]=0$. Using the change of variables
$\alpha \rightarrow 1-\alpha$, we also get
\beq
[D_a ,K] =
 K^b [D_a , D_b] ~+~ {1\over 2} K^{bc} [[D_a, D_b],D_c]~+\cdots
 \label{mat10}
 \eeq
Since the second term on the right hand side is of order $1/n$, and commutators will
further reduce the power of $n$, by adding (\ref{mat9}) and (\ref{mat10}), we find
\beq
[D_a , K] = {1\over 2} \left( \Omega_{ab} K^b + K^b \Omega_{ab} \right) ~+~
{\cal O} (1/n^2)
\label{mat11}
\eeq
Equivalently, we can write this as
\beqar
\Omega_{bc} K^c &=& [D_b, K] ~+~ {1\over 2}  [\Omega_{bc}, K^c] ~+{\cal O}(1/n^2)
\nonumber\\
K^c \Omega_{bc} &=& [D_b, K] ~-~ {1\over 2}  [\Omega_{bc}, K^c] ~+{\cal O}(1/n^2)
\label{mat12}
\eeqar
Using this result, equation (\ref{mat8a}) may be simplified as
\beqar
\Tr ~(\delta K)&=&- {1\over 2} \Tr \Bigl[ \delta D_a N_{ab} [D_b, K] 
- [D_b, K] N_{ba} \delta D_a \nonumber\\
&&\hskip 1in + {1\over 2}
\Bigl( \delta D_a N_{ab} [\Omega_{bc}, K^c] - [\Omega_{bc}, K^c] N_{ab}
\Bigr) + {\cal O}(1/n^3)\Bigr]\nonumber\\
&=&-{1\over 2} \Tr \Bigl[ \delta D_a N_{ab} [D_b, K] 
- [D_b, K] N_{ba} \delta D_a + {1\over 2}
\delta D_a [N_{ab}, [\Omega_{bc}, K^c]]
+ {\cal O}(1/n^3)\Bigr]\nonumber\\
&=&-{1\over 2} \Tr \Bigl[ \delta D_a N_{ab} [D_b, K] 
- [D_b, K] N_{ba} \delta D_a + {\cal O}(1/n^3)\Bigr]\label{mat13}
\eeqar
We have used the fact that $[N_{ab}, [\Omega_{bc}, K^c]]$ is of order
$1/n^3$ since $N_{ab}$ is of order $1/n$.
This result shows that the change in $\Tr~ K$ is given by the first set of terms to
order $1/n^2$, as $n$ becomes large.

\noindent{$\underline{Variation ~of ~ \Tr (D_0)}$}

The expansion of the action (\ref{GM1}) can now be carried out, taking
$K= D_0$. The relevant terms in $N_{ab}$, from equation (\ref{N12}), are
\beq
N_{ab} = \omega^{-1}_{ab}  + {(k-1) \over 4 B_n} \delta_{ab} 
-{\omega^{-1}_{ac}~F_{cd} ~\omega^{-1}_{db} } - {\mathbb R}_{ac} 
~\omega^{-1}_{cb}
~+{\cal O}(1/n^3)
\label{mat14}
\eeq
where
\beq
\omega^{-1}_{ab} = {f_{abc} D_c \over B_n} \label{mat15}
\eeq
(This will become the inverse of the symplectic form in the large $n$ limit; the notation anticipates this. However, at this stage $\omega^{-1}_{ab}$ is still a matrix.)
Upon using this in (\ref{mat13}), the term involving $\delta_{ab}$ will drop out by cyclicity of the trace, the term involving ${\mathbb R}_{ac}$ will turn out to be of order $1/n^3$, as shown 
later.  For the variation of the action, up to order $1/n^2$, we then get
\beqar
\Tr~(\delta D_0) &=& 
\delta D_0^{(1)} + \delta D_0^{(2)} +\cdots\nonumber\\
\delta D_0^{(1)}&=&- {1\over 2} \Bigl( \delta A_a \omega^{-1}_{ab} F_{b 0}
+ F_{b 0} \omega^{-1}_{ab} \delta A_a\Bigr)\label{mat16}\\
\delta D_0^{(2)}&=& {1\over 2} \Bigl(\delta A_a \omega^{-1}_{ab} F_{bc}
\omega^{-1}_{cd} F_{d 0}
+ F_{d 0} \omega^{-1}_{ab} F_{bc} \omega^{-1}_{cd} \delta A_a
\Bigr)\nonumber
\eeqar
This can now be converted to an integral of the symbol of this expression over 
${\bf CP}^k$ with a trace over the remaining (small) matrix labels.
First of all, in $(\delta D_0^{(1)})$, we bring the $\omega^{-1}_{ab}$
to the left end by the cyclicity of the trace, and then we can take it out of the symbol
by using (\ref{res18}).
In other words, for this term, we can write
\beq
\omega^{-1}_{ab} ={1\over B_n}\left[  -i {nk \over \sqrt{2k(k+1)}} f_{abc} S_{c,k^2+2k} 
+ {i\over 2} f_{abc} S_{c-i} R_{+i} + f_{abc} S_{ci} A_i\right]
\label{mat17}
\eeq
where we have also used the fact the symbol of the gauge field may be written
as $A_c = S_{ci}A_i$ where the summation is over $i = 1$ to $2k$.
The fact that the symbol of $A_c$ has this restricted form is due to the constraints
(\ref{GM3}).
The first term on the right hand side of (\ref{mat17}) is also related to
$\omega^{-1\mu\nu}$ in the coordinate basis as
\beqar
-i {nk \over B_n\sqrt{2k(k+1)}}f_{abc} S_{c,k^2+2k} &=&
-i {nk \over B_n\sqrt{2k(k+1)}}~ f^{ij, k^2+2k} S_{ai } S_{bj}\nonumber\\
&=& \omega^{-1\mu\nu} E^i_\mu E^j_\nu S_{ai} S_{bj}\label{mat18}\\
\omega^{-1\mu\nu}&=&-i {nk \over B_n\sqrt{2k(k+1)}}~ f^{ij, k^2+2k} (E^{-1})^\mu_i
(E^{-1})^\nu_j\nonumber
\eeqar
$E$'s are the frame fields for the metric on ${\bf CP}^k$.
There is one more
correction which we must take account of. Originally, we defined the symbol using the states with
the gauge potential equal to zero. As we change the potential, the definition of the symbol also changes. This correction may written, for any matrix $K$, as
\beq
(K) = (K)_0 - {1\over 4}  \bigl( (\omega^{-1}_{ab} F_{ab} + F_{ab} \omega^{-1}_{ab})~K\bigr)
~+ \cdots
\label{mat20}
\eeq
To keep the continuity of our main line of argument, we do not give the derivation of this result here; it is shown in the next subsection. 
For our case of applying this to (\ref{mat16}), we take $K = - {1\over 2}( \delta A_a \omega^{-1}_{ab} F_{b 0}
+ F_{b 0} \omega^{-1}_{ab} \delta A_a )$.
In using equation (\ref{mat20}),  $\omega^{-1}$ is given by just the
first term in (\ref{mat17}), since that is all we need to get terms ot order $1/n^2$.
Similarly, for $(\delta D_0^{(2)})$, we can replace $\omega^{-1}_{ab}$ by the first term 
on the right hand side of (\ref{mat17}).

We can now write
\beqar
\int dt ~\Tr (\delta D_0^{(1)})&=& -{1\over 2} \Tr \Bigl[ \omega^{-1}_{ab} \Bigl(
\delta A_a F_{b0} + F_{b0} \delta A_a \Bigr) \Bigr]\nonumber\\
&=&{\rm Term~ I} +{\rm Term ~II} +{\rm Term~ III}+{\rm Term ~IV}
+{\rm Term~V}
\label{mat21}
\eeqar
In writing this expression in terms of symbols, we will also need
the first term of the star product for $\delta A_a F_{b0} + F_{b0} \delta A_a$; this is indicated
as Term II. Explicitly,
\beqar
{\rm Term ~I}&=& N\int dt d\mu ({\bf CP}^k)~
\Bigl[ -{1\over 2} \omega^{-1\mu\nu} E^i_\mu E^j _\nu 
S_{ai} S_{bj} \tr \Bigl( \delta (A_a) (F_{b 0})+
(F_{b 0}) (\delta A_a)\Bigr)\Bigr]\nonumber\\
&=&N\int dt d\mu ({\bf CP}^k)
\Bigl[-  \omega^{-1\mu\nu} \tr ( \delta A_\mu F_{\nu 0})\Bigr]
\label{mat22}
\eeqar
We have used the components of $A$ and $F$ in the coordinate basis.

For the second term, we need the correction from the star product.
For functions which are built up of sums of products of $D$'s, this is given by
the term $S_{a-i} R_{+i}$ when $D$'s are replaced by
the formula (\ref{res18}). Noting that $\delta F /\delta D$ is given by
$N_{ab} [D_b, F]$ according to (\ref{mat12}), we can write
\beqar
F*G &=& FG + {nk \over 2 \sqrt{2k(k+1)}} S_{a-i} (R_{+i} S_{c,k^2+2k} )(N_{ab}) ([D_b,F] ) (N_{cd}) ([D_d, G])  ~+\cdots\nonumber\\
&=& FG ~-~ {n\over 4} \omega^{-1\mu\nu} \omega^{-1\alpha\beta} E^{-k}_\mu
E^{+k}_\alpha [D_\nu ,F] [D_\beta ,G]~+\cdots
\label{mat23}
\eeqar
The second term can be simplified as follows.
\beqar
{\rm Term ~II} &=&N\int dt d\mu ({\bf CP}^k)~\omega^{-1\mu\nu}
E^i_\mu E^j_\nu S_{ai} S_{bj} \bigl( E^{-k}_\alpha E^{+k}_\gamma +
E^{+k}_\alpha E^{-k}_\gamma\bigr)\nonumber\\
&&\hskip 1.3in \times
\Bigl[{n\over 8} \omega^{-1\alpha\beta}
\omega^{-1\gamma\delta} \tr \bigl([D_\beta , \delta A_a] [D_\delta , F_{b0}] \bigr)\Bigr]\nonumber\\
&=&N\int dt d\mu ({\bf CP}^k)~ {n \over 4} \omega^{-1\mu\nu}
\omega^{-1\alpha\beta}
\omega^{-1\gamma\delta} g_{\alpha\gamma} \tr \bigl([D_\beta , \delta A_\mu] [D_\delta , F_{\nu 0}] \bigr)
\label{mat24}
\eeqar
where $ g_{\alpha\gamma}$ is the metric tensor. The product of
$\omega^{-1}$'s in this equation can be simplified as
\beq
\omega^{-1\alpha\beta} \omega^{-1\gamma\delta} g_{\alpha\gamma}
= - {n^2 \over 4 B_n^2} g^{\beta \delta}
\label{mat25}
\eeq
The expression for the second term now reads
\beqar
{\rm Term ~II}&=& N\int dt d\mu ({\bf CP}^k)\left[ -{n^3 \over 16 B_n^2}~ \omega^{-1\mu\nu}
g^{\alpha\beta}
\tr \Bigl[ (D_\alpha \delta A_\mu ) (D_\beta F_{\nu 0})\Bigr]\right]\nonumber\\
&=&N\int dt d\mu ({\bf CP}^k)\left[ {n^3 \over 16 B_n^2}~ \omega^{-1\mu\nu}
\tr \Bigl[ \delta A_\mu ~D^2~F_{\nu 0} \Bigr]\right]\nonumber\\
&\approx&N\int dt d\mu ({\bf CP}^k)\left[ {1\over n}~ \omega^{-1\mu\nu}
\tr \Bigl[ \delta A_\mu ~D^2~F_{\nu 0} \Bigr]\right]
\label{mat26}
\eeqar

The third term arises from the derivatives $R_{+i}$ in (\ref{mat17}) and reads
\beqar
{\rm Term ~III}&=& N \int dt d\mu ({\bf CP}^k) \left[ -{i\over 2 B_n} f_{abc} S_{c-i} R_{+i}
\tr ( \delta A_a F_{b0}) \right] \nonumber\\
&=& N \int dt d\mu ({\bf CP}^k) \left[ {i\over 2 B_n} f_{abc} (R_{+i}S_{c-i} )~
\tr ( \delta A_a F_{b0}) \right] \nonumber\\
&=& N \int dt d\mu ({\bf CP}^k) \left[ -{k+1\over n} \omega^{-1\mu\nu} \tr (\delta A_\mu F_{\nu 0})
\right]
\label{mat27}
\eeqar
The fourth term involves the combination $f_{abc} S_{ci} \delta A_a F_{b0}$. Since the fields
only involve the coset components, this leads to $f_{ijk}$ where all indices belong to the
coset; this is zero, ${\rm Term~ IV} =0$.

The fifth term in (\ref{mat21}) corresponds to the modification of the symbol due to
the change in $A_a$, as shown in (\ref{mat20}).This is given by
\beqar
{\rm Term ~V} &=& {1\over 4}\Tr \Bigl[\omega^{-1}_{ab} \omega^{-1}_{cd}
\bigl( \delta A_a F_{b0} + F_{b0} \delta A_a\bigr) F_{cd}
\Biggr]\nonumber\\
&=& N \int dt d\mu ({\bf CP}^k) \left[ {1\over 4} \omega^{-1\mu\nu}\omega^{-1\alpha\beta}
\tr \Bigl[( \delta A_\mu F_{\nu 0} + F_{\nu 0} \delta A_\mu ) F_{\alpha \beta} \Bigr]
\right]
\eeqar
Collecting terms from equations (\ref{mat22}, \ref{mat26}, \ref{mat27}), we get
\beqar
\int dt \Tr (\delta D_0^{(1)}) \!\!\!&=&\!\!\! N \int dt d\mu ({\bf CP}^k) \biggl[ -
 \omega^{-1\mu\nu} \tr ( \delta A_\mu F_{\nu 0})\nonumber\\
 &&\hskip .8in +{1\over 4} \omega^{-1\mu\nu}\omega^{-1\alpha\beta}
\tr \Bigl[( \delta A_\mu F_{\nu 0} + F_{\nu 0} \delta A_\mu ) F_{\alpha \beta} \Bigr]\nonumber\\
 &&\hskip 1.2in - {1\over n}~ \omega^{-1\mu\nu}
\tr \Bigl[ \delta A_\mu ~\left( - D^2 + (k+1)\right) ~F_{\nu 0} \Bigr]\biggr]
\label{mat28}
\eeqar

We can simplify $\delta D_0^{(2)}$ in a similar way. Only the leading term in
$\omega^{-1}_{ab}$ will contribute to the order of interest, so the evaluation of this term is
even simpler. Putting it together with (\ref{mat28}), we find
\beqar
\int dt \Tr (\delta D_0 ) \!\!\!&=&\!\!\! N \int dt d\mu ({\bf CP}^k)
\Biggl[-  \omega^{-1\mu\nu} \tr ( \delta A_\mu F_{\nu 0})\nonumber\\
&&\hskip .3in + {1\over 2} \left( \omega^{-1\alpha \mu} \omega^{-1\nu\beta}
+  {1\over 2} \omega^{-1\alpha\beta}\omega^{-1\mu\nu} \right)~\tr 
\Bigl[(\delta A_\alpha F_{\beta 0} + F_{\beta 0} \delta A_{\alpha }) F_{\mu\nu}\Bigr]
\nonumber\\
&&\hskip .3in -{1\over n}~ \omega^{-1\mu\nu}
\tr \Bigl[ \delta A_\mu ~\left(-D^2 +(k+1)\right)~F_{\nu 0} \Bigr] +{\cal O}(1/n^3)\Biggr]
\label{mat29}
\eeqar

\noindent{$\underline{Relation~ to ~the~ Chern-Simons ~action}$}

The first set of terms in the expression (\ref{mat29}) can be related to the Chern-Simons form as before.
The variation of the $(2k+1)$-dimensional Chern-Simons term gives
\beq
\delta S = {i^{k+1} \over (2\pi )^{k} k!} \int \tr \bigl( \delta A F^k \bigr)
\label{mat30}
\eeq
We replace $F$ by $\omega +F$ and write this out as
\beq
\delta S = {i^{k+1} \over (2\pi )^{k} k!} \int \tr \Bigl( \omega^k \delta A  + k \omega^{k-1} 
\delta A F + \half k (k-1)  \omega^{k-2} \delta A F^2 +\cdots \Bigr)
\label{mat31}
\eeq
(Various terms in this expansion are obviously related to K\"ahler-Chern-Simons actions
\cite{nair2}.) Equation (\ref{mat31}) can be simplified using equations (\ref{ab18}, \ref{ab18a}). The first term is
\beq
{i^{k+1} \over (2\pi )^{k} k!}~\omega^k \tr (\delta A) = i {n^k \over k!}dt d\mu ({\bf CP}^k) ~ \tr (\delta A_0) 
\label{mat32}
\eeq
The second term can be simplified as
\beqar
{i^{k+1} \over (2\pi )^{k} k!}~k \omega^{k-1} \tr (\delta A_\mu F_{\nu 0} )~ dx^\mu  dx^\nu  dt
\!\!\! &=&\!\!\! i^{k+1} k {2^k k!\sqrt{\det \omega} \over (2\pi )^k 2^{k-1}} d^{2k}x dt ~\left[ - {1\over 2k}\omega^{-1\mu\nu}\right]
\tr (\delta A_\mu F_{\nu 0} )\nonumber\\
\!\!\! &=&\!\!\! i {n^k \over k!} \int dt d\mu ({\bf CP}^k) ~ \Bigl[ - \omega^{-1\mu\nu} \tr (\delta A_\mu F_{\nu 0})\Bigr]
\label{mat33}
\eeqar
For the third term we get
\beqar
{i^{k+1} \over (2\pi )^{k} k!} {k (k-1)\over 2} \omega^{k-2} \tr ( \delta A F F )
&=& i {n^k \over k!} \int dt d\mu ({\bf CP}^k) ~\half\left[ \omega^{-1\mu\alpha} \omega^{-1\beta\nu}
+ \half \omega^{-1\mu\nu} \omega^{-1\alpha\beta}\right]\nonumber\\
&&\hskip 1in \times \tr \Bigl[ (\delta A_\alpha F_{\beta 0}
+ F_{\beta 0} \delta A_\alpha ) F_{\mu\nu} \bigr]
\label{mat34}
\eeqar
Comparing (\ref{mat33}) and (\ref{mat34}) with (\ref{mat29}), we see that
\beqar
i \int dt~\Tr (\delta D_0) &=& {N k!\over n^k} {i^{k+1} \over (2\pi )^{k} k!}  \left[
k \omega^{k-1} \tr (\delta A F) + \half k (k-1) \omega^{k-2}
\tr (\delta A FF) +\cdots \right]\nonumber\\
&&\hskip .0in -i{N\over n} \int dt d\mu ({\bf CP}^k)~ \omega^{-1\mu\nu}
\tr \Bigl[ \delta A_\mu ~\left(-D^2 +(k+1)\right)~F_{\nu 0} \Bigr] +{\cal O}(1/n^3)\nonumber\\
\label{mat35}
\eeqar
This gives the change in the expression for $i \Tr ( D_0 )$ as we expand around a background with potential
$A_\mu +\delta A_\mu$ rather than $A_\mu$. There could also be a change due to
variation of $A_0$, i.e., due to a change in the functional form of $A_0$, which is given by
\beq
i \Tr \delta_f A_0 = {N k! \over n^k} {i^{k+1} \over (2\pi )^{k} k!} 
\omega^k \tr (\delta A_0)
\label{mat36}
\eeq
Including this, we find, for the total variation,
\beqar
i \int dt~\Tr (\delta D_0) \!\!\!&=&\!\!\! {N k! \over n^k}~\delta S_{CS} 
-i{N\over n} \int dt d\mu ({\bf CP}^k)~ \omega^{-1\mu\nu}
\tr \Bigl[ \delta A_\mu \left(-D^2 +(k+1)\right)F_{\nu 0} \Bigr]+\cdots\nonumber\\
\!\!\! &\approx&\!\!\!  \delta S_{CS} -i{N\over n} \int dt d\mu ({\bf CP}^k)~ \omega^{-1\mu\nu}
\tr \Bigl[ \delta A_\mu \left(-D^2 +(k+1)\right) F_{\nu 0} \Bigr]+{\cal O}(1/n^3)\nonumber\\
\label{mat37}
\eeqar
The first term on the right hand side of this equation integrates to give the Chern-Simons form. The second term is due to the higher terms in the star product. This is clearly seen from
the fact they arise from the higher terms in (\ref{mat23}) and from the
term involving $R_{+i}$ in the large $n$ limit of $N_{ab}$ In fact, we may write the variation of
the action in (\ref{mat29}) as
\beqar
i \int dt \Tr (\delta D_0 ) \!\!\!&=&\!\!\! i N \int dt d\mu ({\bf CP}^k)
\Biggl[-  \omega^{-1\mu\nu} *\tr ( \delta A_\mu* F_{\nu 0})\nonumber\\
&&\!\!\!\!\!\! + {1\over 2} \left( \omega^{-1\alpha \mu} \omega^{-1\nu\beta}
+  {1\over 2} \omega^{-1\alpha\beta}\omega^{-1\mu\nu} \right)\tr 
\Bigl[(\delta A_\alpha F_{\beta 0} + F_{\beta 0} \delta A_{\alpha }) F_{\mu\nu}\Bigr]
+{\cal O}(1/n^3)\Biggr]\nonumber\\
\label{mat38}
\eeqar
We could also write the second set of terms with star products between the various 
factors, since, to the order we have calculated, they would just become ordinary products.
This means that the right hand side can be interpreted as the variation of 
$S_{*CS}$, the Chern-Simons term defined with star products used for the products of
fields and their derivatives occurring in it.
The integrated version is thus
\beq
i\int dt~ \Tr (D_0) \approx S_{*CS}~+~\cdots
\label{mat39}
\eeq

\noindent{$\underline{Simplification~ in~ a ~gradient ~expansion}$}

Actually, we can go a little further. We define the basic scale parameter, the analog of radius
for the manifold, by $n =2 BR^2$, $B$ being the value of the `magnetic field'.
The prefactor for the action scales as $N \sim n^k \sim R^{2k}$ and this can be absorbed into the definition of the volume element for the space. In other words, the volume we have defined is in terms of dimensionless coordinates, so we write
$R^{2k} d\mu ({\bf CP}^k) = dV$. $\omega^{-1\mu\nu}$ arises from
$f_{abc} D_c /B_n$ and so it has the form
\beqar
\omega^{-1\mu\nu}&\sim& {1\over R^2} f^{ij, k^2+2k} (E^{-1})^\mu_i
(E^{-1})^\nu_j\nonumber\\
&\equiv&{1\over R^2} {\tilde \omega^{-1\mu\nu}}
\label{mat40}
\eeqar
Since $A_\mu$ must scale like the derivative, we can define the gauge fields of proper dimension as ${\tilde A}_\mu  = A_\mu /R$. Thus the combination $\omega^{-1\mu\nu}
\tr (\delta A_\mu F_{\nu 0})$ is independent of $R$, written in terms of ${\tilde A}_\mu$ as
${\tilde \omega^{-1\mu\nu}}
\tr (\delta {\tilde A}_\mu {\tilde F_{\nu 0}})$. The term in (\ref{mat38}) with
two powers of $\omega^{-1}$ is also independent of $R$.  Among the terms
in (\ref{mat37}) which arise from the star product, we find
\beq
-i {N\over n}(k+1) \int dt d\mu ({\bf CP}^k) \omega^{-1\mu\nu}
\tr \Bigl[ \delta A_\mu F_{\nu 0}\Bigr]
= {k+1\over 2 BR^2} \int dV ~\tr \Bigl[ \delta {\tilde A}_\mu {\tilde F}_{\nu 0}\Bigr]
\label{mat41}
\eeq
This term is therefore negligible in the large $R$ limit.
The remaining term in (\ref{mat37}), namely, the term involving $(-D^2)$, becomes
\beq
-i {N\over n} \int dt d\mu ({\bf CP}^k) \omega^{-1\mu\nu}
\tr \Bigl[ \delta A_\mu (-D^2) F_{\nu 0}\Bigr]
= - {i\over 2 B} \int dt dV~ \tr \Bigl[ \delta {\tilde A}_\mu
(-{\tilde D}^2) {\tilde F}_{\nu 0}\Bigr]
\label{mat42}
\eeq
This term is evidently independent of $R$, and so, survives the large $R$-limit.
However, notice that it involves the spatial gradients of the field $F_{\nu 0}$. We can thus envisage a gradient expansion where we keep $F_{\nu 0},~F_{\mu\nu}$, as in the Chern-Simons term, but consider gradients of these fields to be smaller, i.e.,
$\vert D^2 F_{\nu 0}\vert \ll \vert F_{\mu\nu}\vert \vert \omega^{-1}\vert \vert F_{\nu 0}\vert$.
In this case, the term (\ref{mat42}) is also small.

Thus, if we take the large $R$ limit, and consider the leading terms in a gradient expansion,
the result of our long analysis is
\beq
i\int dt~ \Tr (D_0) \approx S_{CS}~+~\cdots
\label{mat43}
\eeq
This is one of the main results of this paper. It shows that the expansion of
$\Tr (D_0)$ around different backgrounds can be approximated, in the large $n$ limit
and for small gradients for the field strengths,
by the Chern-Simons form, with $A$ replaced by $a+A$, $a$ being the desired background
potential.

Strictly speaking, our calculation has explicitly verified this result (\ref{mat43}) only up to
order $1/n^2$, or, equivalently, up to the term involving the $5$-dimensional
Chern-Simons form for $A$. To this order, we do get $S_{CS}(a+A)$.
The full result, whatever it is, should be a functional of only the combination  $a+A$, since the separation between the background and fluctuation is arbitrary, it should have the correct gauge invariance property and it should agree with $S_{CS}(a+A)$ when expanded up to the term with the
$5$-dimensional Chern-Simons form for $A$. The only such term, apart from the ambiguity of higher
gradients of fields, is $S_{CS}(a+A)$. This argument completes the demonstration of
our result (\ref{mat43}).

\subsection{Two ancillary results}

We shall now show two results which were
needed for the analysis done in the last subsection. 
We will begin with the result (\ref{mat20}) which is for the change in the symbol for a matrix.

\noindent{$\underline{Change~ in~ the~ symbol~ for~ a~ matrix}$}

We go back to (\ref{mat13}) and write the first correction to 
$\Tr K$ due to the change in $A_a$ as
\beq
\delta_A \Tr (K) = - {1\over 2} \Tr \Bigl[ \delta D_a N_{ab} [D_b, K] 
- [D_b, K] N_{ba} \delta D_a\Bigr]
\label{sym1}
\eeq
If we also consider a change in the functional form of $K$, denoted by
$\delta_f K$, we get
\beqar
\delta_f \delta_A \Tr K &=& - {1\over 2} \Tr \Bigl[ \delta D_a N_{ab} [D_b, \delta_f K] 
- [D_b, \delta_f K] N_{ba} \delta D_a\Bigr]\nonumber\\
&=&{1\over 2} \Tr \Bigl[ [D_b, \delta D_a N_{ab} - N_{ba} \delta D_a ] ~\delta_f K\Bigr]
\label{sym2}
\eeqar
Here we are interested in the term of order $1/n$. For this, we can take $N_{ab} = \omega^{-1}_{ab} = f_{abc} D_c /B_n$ and write
\beqar
[D_b , \delta D_a N_{ab} ]&=& {f_{abc} \over B_n}\left( [D_b , \delta D_a ] D_c +
\delta D_a [D_b , D_c]\right)\nonumber\\
&=&{f_{abc} \over B_n}\left( {1\over 2} \delta ([D_b , D_a ])~ D_c +
\delta D_a ( f_{bck} D_k +F_{bc} )\right)\nonumber\\
&=& \left[ {k+1 \over 2 B_n} \delta D \cdot D - {1\over 2}\delta F_{ab}~ \omega^{-1}_{ab}
+ {f_{abc} \delta D_a F_{bc}  \over B_n}\right]
\label{sym3}
\eeqar
In a similar way,
\beq
- [D_b , N_{ba} \delta D_a ] =
\left[ {k+1\over 2 B_n}  D \cdot \delta D - {1\over 2} \omega^{-1}_{ab} \delta F_{ab}
+ {f_{abc} F_{bc} \delta D_a \over B_n}\right]
\label{sym4}
\eeq
The symbol corresponding to $f_{abc} F_{bc} \delta D_a$, which occurs in the above expression, is of the form $f_{abc} S_{ai} S_{bj} S_{ck} F_{ij} \delta A_k$,
where $i, j ,k $ are coset indices, 
since the fields have only coset components at the symbol-level; this combination is then zero.
Further, using $\delta D^2 =0$, we find
\beq
\delta_f \delta_A \Tr K = - {1\over 4} \Tr \left( \delta F_{ab} \omega^{-1}_{ab} +
\omega^{-1}_{ab} \delta F_{ab} \right)
\label{sym5}
\eeq
In $\omega^{-1}_{ab} = f_{abc} (-iT_c +A_c)/B_n$, the second term is of order
$1/n^2$. Keeping this in mind, we may integrate over $A$ to obtain
\beq
\Tr \delta_f K = \Tr (\delta_F K)\Bigr]_{A=0} ~-~ {1\over 4} \Tr \Bigl[(F_{ab} \omega^{-1}_{ab}
+\omega^{-1}_{ab} F_{ab} ) \delta_f K \Bigr] ~+~\cdots
\label{sym6}
\eeq
(On the right hand side, it is sufficient to retain the $f_{abc}(-iT_c)$ part of
$\omega^{-1}_{ab}$.)
This equation shows that there is a modification of the symbol to be used in the presence of nonzero $A_a$, given for any matrix $K$ as
\beq
(K) = (K)_0 - {1\over 4} \bigl( ( F_{ab}\omega^{-1}_{ab} + \omega^{-1}_{ab} F_{ab}) ~K\bigr)
~+\cdots
\label{sym7}
\eeq
This is the same as equation (\ref{mat20}).

\noindent{$\underline{The ~contribution ~of ~{\mathbb R}_{ab}}$}

We shall now show that the contribution of ${\mathbb R}_{ab}$ is of order
$1/n^3$ to the variation of the action. 
We are interested in the term of order $n$ in the expression for $R_{ab}$, which was
defined in equation (\ref{N1}). These terms are given by
\beq
R_{ab} \approx -i f_{ack} f_{cbl} (T_k A_l + A_k T_l )
+i (T_b A_a + A_b T_a) - i \left( n + {k+1\over 2}\right) d_{abc} A_c + {\cal O}(1)
\label{rab1}
\eeq
Using the formula (\ref{N3}) for the product of $f$'s, this can be written as
\beqar
i R_{ab} &\approx& - {2\over k+1} (T\cdot A + A\cdot T)\delta_{ab}  - {k-1\over k+1}
(T_b A_a +A_b T_a)+\left( n + {k+1\over 2}\right) d_{abc} A_c\nonumber\\
&&\hskip .2in - 4 ( d_{abm} d_{klm}-  d_{alm} d_{kbm}) (T_k A_l +T_l A_k) 
 + {\cal O}(1)
\label{rab2}
\eeqar
$T\cdot A + A\cdot T \approx 0$ to this order, since $D^2 = -C_n$. Further, the second condition in (\ref{GM3}) gives 
\beq
d_{klm} (T_l A_m + A_m T_l)\approx {k-1 \over 4(k+1)}\left( n +{k+1 \over 2}\right) ~A_k
\label{rab3}
\eeq
Using these, we can simplify (\ref{rab2}) as
\beq
i R_{ab} \approx - {k-1\over k+1} (T_b A_a +A_b T_a ) + {2n + k+1 \over k+1} d_{abc} A_c
+ 4 d_{alm} d_{kbm} (T_k A_l + A_k T_l ) +{\cal O}(1)\label{rab4}
\eeq
We now take the symbol of this expression. The term with the $d$-symbol becomes, for example,
\beq
(d_{kbm} T_k ) \approx d_{kbm} {nk \over \sqrt{2k (k+1)}} S_{k, w}
= {nk \over \sqrt{2k (k+1)}} d_{w\alpha \beta} S_{b,\alpha} S_{m,\beta} 
\label{rab5}
\eeq
where, for brevity, we use the index $w$ to denote the $(k^2+2k)$-th component.
For the $d$-symbol we have the formula
\beq
d_{w\alpha \beta} = \left\{ \begin{array}{l c l} - {1\over \sqrt{2k(k+1)}} \delta_{\alpha \beta}
&~&\alpha, \beta \in \underline{SU(k)}\\
{k-1\over 2}{1\over \sqrt{2k(k+1)}}\delta_{\alpha \beta}&~ &\alpha ,\beta \in \underline{SU(k+1)}
-\underline{U(k)}\\
{k-1\over 2}{1\over \sqrt{2k(k+1)}}&~& \alpha =\beta = w\\
\end{array}
\right.
\label{rab6}
\eeq
(The signs are as given for $t_a = -\half \lambda^T_a$.) Equation (\ref{rab5})
can be worked out as
\beq
(d_{kbm} T_k ) \approx - {n \over 2 (k+1)}\delta_{bm}  + {n\over 4}
(S_{b i} S_{mi}  + S_{bw} S_{mw})
\label{rab7}
\eeq
The gauge potential can be taken to have the form $A_l = S_{li} A_i$ where 
$i$ is a coset index taking values $1, 2, ..., 2k$. The symbol for $R_{ab}$ is now simplified
as
\beqar
i R_{ab} &\approx& A_i \biggl[ - {k-1\over k+1} {nk \over \sqrt{2k(k+1)}} (S_{bw}S_{ai}
+ S_{bi} S_{aw} ) + {n (k-1) \over 2 \sqrt{2k(k+1)}}  (S_{bw}S_{ai}
+ S_{bi} S_{aw} )\nonumber\\
&&\hskip .5in - {2n \over k+1} d_{abc} A_c\biggr]
+ n ~d_{\alpha ij} \left[ S_{a \alpha} S_{bi} A_j + S_{b\alpha} S_{ai} A_j\right]+
{\cal O}(1)
\label{rab8}
\eeqar
The index $\alpha$ in the last term on the right hand side can be $w$ or 
an $SU(k)$ index. Again writing $A_c = S_{ci} A_i$ and collecting similar terms, we
find
\beq
iR_{ab} \approx n {k-1 \over k+1} d_{Aij}
(S_{aA}S_{bi} + S_{ai} S_{bA} ) A_j +{\cal O}(1)
\label{rab9}
\eeq
where the index $A$ refers to the $SU(k)$-algebra. This contributes a term of
order $1/n$ in ${\mathbb R}_{ab} = R_{ab}/B_n$. However, it does not contribute
to the variation of $\Tr D_0$ at this order. The relevant term in
the variation of the action comes from the term
$({\mathbb R} N_0 )_{ab}$ in the expansion of $N_{ab}$ as in (\ref{N12})
and is given by
\beqar
i \delta \Tr D_0 &\approx&  \Tr (\delta A_a {\mathbb R}_{ab} N_{0bc} F_{c0} )
~+~h.c.\nonumber\\
&\approx& -i {n^2 k (k-1)\over B^2_n\sqrt{ 2k(k+1)^3}}d_{Ajk}
\delta A_i S_{ai} (S_{aA} S_{bj} + S_{aj} S_{bA} )
f_{wrs} A_k S_{br} S_{cs}~+~ h.c.\nonumber\\
&=&0
\label{rab10}
\eeqar
where the indices $i,j,k,r,s$ are in the coset and we have used the fact that
$S_{ai} S_{aA}=0$, $S_{bA} S_{br}=0$. Thus ${\mathbb R}_{ab}$ does not contribute to this
order (i.e., at $1/n^2$ order) to the variation of $\Tr D_0$. Its contribution starts at
order $1/n^3$.

\section{Applications}

\subsection{Quantum Hall effect in arbitrary dimensions}

The analysis we have done can be applied to some physical situations.
The first and simplest case is to the quantum Hall effect \cite{ZH, KN, other}. When Hall effect is generalized to higher dimensions, generically, we will be dealing with ${\bf CP}^k = SU(k+1) /U(k)$.
The ``constant'' background magnetic field then takes values in the Lie algebra of $U(k)$, the local isotropy group of the space. It is then possible to consider fluctuations $\delta A$ in these gauge fields. The effective action for these fluctuations will have two contributions, the bulk part and the boundary part, the later describing the boundary excitations of a droplet of
fermions. The bulk part may be viewed as the integration of $\int J\cdot \delta A$ where
$J$ is the Hall current. It is clear that our calculation will
give this bulk action as $S_{CS}(a+A)$.

As mentioned in the introduction, this has been calculated in reference \cite{karabali}
starting from the matrix action $S =i\int dt \Tr (U^\dagger D_0 U)$. The gauge invariance of this action requires the transformation of the gauge potential as given by
$D_0 \rightarrow e^{-i\lambda} D_0 e^{i\lambda}$, where $\lambda$ is some hermitian matrix.
On the other hand, in the large $n$ limit, we have background gauge fields $A_i$
which are functions on the space ${\cal M}$ and transform as $A_i \rightarrow
e^{-i\Lambda} (\del_i +A_i )e^{i\Lambda}$, where $\Lambda$ is a $M\times M$-matrix-valued function. The idea is to regard the symbol of $A_0$ as a function of $A_i$ so chosen that
the $\Lambda$-transformation of $A_i$ induces the $\lambda$-transformation of
$A_0$ at the level of symbols. One can solve for $A_0$,
and the bulk action $\int dt~\Tr (D_0)$, as series in powers of
$1/n$. The calculation also requires a background potential $V$ to confine the fermions to a droplet, which may be taken as a part of $A_0$, namely, by writing $A_0 \rightarrow A_0 +V$.  
The final result in \cite{karabali} is then
\beq
S = i\int dt~ \Tr (D_0) = S_{CS} (A_0 +V, a+A)
\label{qhe1}
\eeq
Evidently, this is identical to our result, although our method has been 
along the lines of extracting different large $n$ limits for the matrix one-dimensional Chern-Simons action.

\subsection{A matrix version of gravity}

Gauge fields in the case of quantum Hall effect take values in the Lie algebra of $U(k)$.
From the matrix model point of view, where we write matrices in terms of the ${\cal H}_N
\otimes {\cal H}_M$ splitting, there is no obstruction to extending this to $SU(k+1)$ or
even any unitary group. (If we allow all possible types of fluctuations, it is a unitary group
$U(l)$ for some $l$, rather than $SU(l)$, that is relevant.)
At this stage, it is worth recalling that ordinary Minkowski space may be considered as
the coset space $P/L$, where $P$ is the Poincar\'e group and $L$ is the Lorentz group. For the case of ${\bf CP}^k = SU(k+1)/U(k)$, the group $SU(k+1)$ is the analog of the Poincar\'e group
and the isotropy group $U(k)$ is the analog of the Lorentz group. The gauge fields we have introduced correspond to gauging of these groups, and so,  the most natural interpretation
for them is in terms of gravitational degrees of freedom. In other words, our framework and action naturally lead to a matrix version of gravity. (There are many points of connection
between quantum Hall droplets and gravity
\cite{beren}. Our use of the quantum Hall analogy is different;
we focus on constructing gravity on a fuzzy space.)

There are, evidently, some missing ingredients which have to be taken care of before
this can be interpreted in terms of gravity.
First of all, the gauge fields are of the form $A^a_\mu dx^\mu (-it^a)$, which are one-forms on
${\bf CP}^k \times {\bf R}$ (because $dt$ is included in this expression) and the Lie algebra matrices form a basis for $U(k+1)$, or some other unitary group. There is, so far, no analog
of $e^0_\mu dx^\mu$ or $\omega^{0a}_\mu dx^\mu$, corresponding to the time-components of the frame field or spin connection.
Secondly, there is a dimensional mismatch since $U(k+1)$ gives $(k+1)^2$ one-form fields on a $(2k+1)$-dimensional space, where as we we need $(2k+1)(k+1)$ one-forms to describe gravity.
Thirdly, the matrix traces are naturally positive and lead to Euclidean signature for the tangent space. 
We shall now discuss some of these questions, leaving a more detailed treatment for
a separate article \cite{nair}.

A fuzzy space is described by a finite dimensional Hilbert space. This means that 
points on the space are described by states of a Hilbert space. Our point of view is that
these degrees of freedom should therefore be treated exactly as the states of any quantum system. The choice of states in any quantum system is given as density matrix
$\rho_0$, with its evolution given by
\beq
i \del_0 \rho = i {\del \rho \over \del t} = [K, \rho ]
\label{grav1}
\eeq
This equation is essentially the definition of $K$. However, if $K$ is given, it can be taken as defining the evolution of $\rho$.
The action which leads to this equation is given by
\beq
S = i \int d t~ \Tr \bigl[\rho_0~ U^\dagger (\del_0 + A_0 ) U \bigr]
\equiv i \int dt~ \Tr \bigl[\rho_0~ U^\dagger D_0 U \bigr]
\label{grav2}
\eeq
where $A_\tau = i K$. 
This action (\ref{grav2}) has a natural gauge invariance,
$U\rightarrow M U$, $ A_0 \rightarrow M A_0 M^\dagger - \del_0 M M^\dagger$.
The action (\ref{grav2}) is to be regarded as a function of $U$ and gives equation (\ref{grav1}) as the variational equation $\delta S =0$ for variations of $U$.

Our strategy is then the following.
We separate this quantum system into a 
part corresponding to the degrees of freedom of space and a part which describes all other, material, degrees of freedom,
labeling the states as
$\vert \alpha , A\ra$.
The Greek labels $\alpha, \beta,$ etc., pertain to the degrees of freedom of space
and the labels $A, B,$ etc., describe the material system of interest. 
Correspondingly, the operator $D_0$ is separated as
\beq
\la \alpha , A\vert D_0 \vert \beta ,B\ra
=  \delta_{AB} ~\la \alpha \vert D^{(e)}_{0}\vert \beta\ra ~+~ \la \alpha, A\vert D^{(s)}_0
\vert \beta, B\ra
\label{grav3}
\eeq
The operator $D^{(e)}_0$ is relevant for the dynamics of
space and the remainder, $D^{(s)}_0$, describes the material part.
The latter includes effects
of coupling the material system to the spatial degrees of freedom.
Since all the spatial points are to be included in our description, the
density matrix is the identity for the space-part,
\beq
\la \alpha , A\vert \rho_0 \vert \beta , B\ra = \delta_{\alpha \beta} ~\la A \vert \rho_0 \vert B\ra
\label{grav4}
\eeq

Our proposal for fuzzy gravity is then the following. We take the action (\ref{grav2})
as the action for the theory, including gravity, where $U$ and $D^{(e)}_0$ are regarded as 
quantities to be varied. $D^{(s)}_0$ is to be regarded as a given operator, specifying the subsystem of interest. 

The notion of continuous space emerges in the limit of the 
dimension of ${\cal H}_{\cal N}$ becoming large. One may regard 
${\cal H}_{\cal N}$ as arising from the quantization of some phase space ${\cal M}$,
with an appropriate symplectic form. The background fields on this phase space can be varied.
Thus it is possible to calculate the action, expanding $D^{(e)}_0$ in terms of 
the background gauge fields, in the limit of the dimension of ${\cal H}_{\cal N}$ becoming large. This is, of course, what we have done in the previous sections. The best background to expand around is then given by the extremization of the action. 

If we ignore the matter degrees of freedom and concentrate just on the space-part,
then we can approximate
the action as
\beq
S \approx i \int dt~ \Tr ( D^{(e)}_0 ) 
\label{grav5}
\eeq
For expansion around ${\bf CP}^k$, we can simplify this, in the large $n$ limit and for slowly varying fields, as
the Chern-Simons action $S_{CS}$ with the gauge fields being
$a +A$. Notice that this depends only on the full gauge field, the separation into
$a$ and the fluctuation $A$ is immaterial. Renaming the combination $a+A$ as $A$,
we see that the choice of spatial geometry is determined by the CS action \cite{zanelli}.
The isometries of ${\bf CP}^k$ correspond to $SU(k+1)$, so that the
natural choice for the gauge group which occurs in the CS action, i.e., $U(M)$, is
$U(k+1)$. 

It is possible to handle the dimensional mismatch by using an idea similar to compactification. The simplest case is for $k=3$, corresponding to the group $U(4)$.
The Chern-Simons action is thus defined on a seven-dimensional space, which we assume has the topology
$S^2 \times M^5$, where $M^5$ is some five-dimensional manifold.
Writing $U(4) \sim SU(4) \times U(1)$, the gauge field is taken to be of the form
$-i l \omega_K + {\cal F}$, where $\omega_K$ is the K\"ahler form of the two-sphere
$S^2$, $l$ is an integer and ${\cal F}$ belongs to the $SU(4)$ Lie algebra. The effective action is then given by the level $l$, five-dimensional Chern-Simons action with the gauge group $SU(4)$,
\beq
S = -i {l \over 24 \pi^2} \int \tr\left( A ~dA~ dA + {3\over 2} A^3~ dA +{3\over 5} A^5\right)
\label{grav7}
\eeq
Since $SU(4)$ is locally isomorphic to $O(6)$, we see that we have the correct set of gauge fields to describe Euclidean gravity in five dimensions.
In fact,  the gauge potential can be expanded as
\beq
A = P^a ~e^a_\mu dx^\mu ~+~ \half  J^{ab} ~\omega^{ab}_\mu dx^\mu
\label{grav8}
\eeq
where $J_{ab}$ are the generators of $O(5) \subset O(6)$ and
$P_a$ are a basis for the complement of $\underline{O(5)}$ in $\underline{O(6)}$.
$e^a$ are the frame fields and $\omega^{ab}$ is the spin connection.
The variation of the action (\ref{grav7}) can now be simplified as
\beq
\delta S =  - {l\over 128\pi^2} \int~\left[ \delta \omega^{ab} ~{\mathcal R}^{cd}~ 
({\mathcal D} e)^e
+ \delta e^a~ {\mathcal R}^{bc}~ {\mathcal R}^{de}\right]
\epsilon_{abcde}
\label{grav9}
\eeq
where $ ({\mathcal D} e)^a \equiv d e^a +\omega^{ac}e^c$ is the torsion and 
$R^{ab}= d\omega^{ab} + \omega^{ac} \omega^{cb}$ is the Riemann tensor
for the spin connection $\omega^{ab}$. Further, $
{\mathcal R}^{ab} = R^{ab} - e^a ~e^b$.
The equations of motion for gravity, with no matter field, are then
\beqar
\epsilon_{abcde}~ {\mathcal R}^{cd}~ ({\mathcal D} e)^e &=& 0\nonumber\\
\epsilon_{abcde}~ {\mathcal R}^{bc}~ {\mathcal R}^{de}  &=& 0\label{grav11}
\eeqar
The solution to these equations, corresponding to empty space with no matter,
is thus given by
\beq
A = g^{-1} d g, \hskip .5in g \in O(6)
\label{grav12}
\eeq
This space is $O(6)/O(5) =S^5$ which is the Euclidean version of de Sitter space.
It is given in a basis where the
cosmological constant has been scaled out; it may be introduced
by the replacement
$e^a \rightarrow \sqrt{\Lambda} ~e^a$.

\vskip .2in\noindent
I thank Dimitra Karabali for many useful discussions.
I also thank Abhishek Agarwal and H. Steinacker
for useful remarks and for bringing a number of references to my attention.
This work
was supported in part by the National Science Foundation grant number
PHY-0244873 and by a PSC-CUNY grant.

\end{document}